\documentclass[floats,twocolumn,aps,eqsecnum,nofootinbib]{revtex4}

\global\arraycolsep=2pt         % Fix space around '=' in eqnarray's

\usepackage{wrapfig}
\usepackage{graphicx} % Include figure files
\usepackage{dcolumn}  % Align table columns on decimal point
\usepackage{bm}       % Bold math: $\bm{\alpha}$
\usepackage{latexsym} % Several additional symbols
\usepackage{fancyhdr} % Fancy header package
\usepackage{amsmath} % Fancy header package

\newcommand{\smA}{{\rm\scriptscriptstyle A}}

\begin{document}

\preprint{\hfill LA-UR-12-25904}

\title{Field Theory of the $d+t \to n + \alpha$ Reaction
\\
Dominated by a $^5{\rm He}^*$ Unstable Particle
}

\author{Lowell S. Brown and Gerald M. Hale}

\affiliation{
Los Alamos National Laboratory
\\
Los Alamos, New Mexico 87545 
\\}

\begin{abstract}

An effective, non-relativistic field theory for low-energy
$dt \leftrightarrow n\alpha$ reaction is presented. The theory
assumes that the reaction is dominated by an intermediate $^5{\rm He}^*$ 
unstable spin $3/2^+$ resonance.  It involves two
parameters in the coupling of the $dt$ and $n\alpha$ particles to the
unstable resonant state, and the resonance energy level --- only
three real parameters in all. All Coulomb corrections to this
process are computed. The resultant field theory is exactly solvable
and provides an excellent description of the $dt$ fusion process.

\end{abstract}

\maketitle

\section{Introduction and Summary}

\subsubsection{Motivation, Purpose}

In this paper, we examine the reaction
\begin{equation} 
d + t \leftrightarrow n + \alpha \,.
\end{equation}
from an effective field theory point of view. We employ modern
techniques of many-body, non-relativistic quantum field theory\footnote{These 
methods are explained in detail in, for example,
the first two chapters of ref.~\cite{Brown}.} to describe this reaction, 
and also make use of the
contemporary ideas of effective field theory.  In the modern effective
field theory approach, stable nuclei (which are treated as particles)
and resonant nuclear states (which are treated as unstable particles)
 are described by individual fields.  The fields that
correspond to asymptotic states produce particles when they act on the
vacuum (no-particle) state. But fields that correspond to resonances
have no corresponding single-particle states\footnote{Fields 
describing unstable, resonance particles are described at
some length in Section 6.3 of Ref.~\cite{Brown}.}. For the
reaction that we consider in this paper, we shall assume that only a
single intermediate resonant state, corresponding to a spin ${3/2}^+$
${^5{\rm He}}^*$, is needed.  Thus we shall have creation and annihilation
fields for this unstable intermediate resonance as well as such fields
for deuteron, triton, alpha, and neutron particles.

The effective field theory is a generalization of the pseudo-potential
method introduced by Fermi \cite{Fermi} in 1936 for low-energy neutron
scattering on molecules. Fermi used a $\delta({\bf r})$ function
potential taken in first Born approximation. The constant multiplying
the $\delta$ function was chosen to give the correct scattering length
on a nucleus.  The use of a field to describe a composite nucleus was
done as early as 1973 by Schwinger \cite{Julian} 
when he described the deuteron
and used this description to re-derive the effective range
formula for the deuteron's form factor and for its photo-disintegration
cross section.  The modern use of field theory methods in nuclear
physics was advocated by Weinberg \cite{Steve} in 1990.

The effective field theory may be viewed as the simplest mathematical
method to implement a ``black box'' description of nuclear reactions
at low energies. This is a theoretical description that uses the
fewest number of parameters. If the process involves a resonant
intermediate state, then an unstable field is needed in addition to
the fields that describe the propagation of the initial and final
particles.  As the energy of the reacting particles is increased,
additional parameters must be included that correspond to coupling
constants for field interactions involving spatial gradients, which
correspond to
interactions that give higher momentum dependence in the reaction
amplitudes. The number of parameters required increases rapidly with
increasing energy.

Here we are concerned with reactions in the low-energy limit, but with
a resonant intermediate state, the $^5{\rm He}^*$ state.  This
introduces three parameters: two constants $g_{dt}$ and $g_{n\alpha}$
for the coupling of the $dt$ and $n\alpha$ fields to the unstable
$^5{\rm He}^*$ field, and the resonant energy of this unstable
field.  

A traditional method to compute coupled channel nuclear reactions is
to use $R$-matrix theory.  This theory entails nuclear channel radii as
well as excited state energies and channel couplings.  The subsequent
companion paper\cite{BHP} describes the one-level $R$-matrix theory for the two
$dt$ and $n\alpha$ channels. This paper explains in detail how the zero
channel radii limit of this $R$-matrix theory is precisely the result
(\ref{ourresult}) below for the effective field theory with the
coupling to the unstable $^5{\rm He}^*$ particle.

There are several motivations for this work. It provides a detailed
example of how effective field theory methods work for a non-trivial
example that involves a higher spin unstable field. Coulomb
corrections appear not only in the initial state, but also in the
unstable field's self energy involving the $dt$ loop. Field theory
methods of angular momentum coupling simplify the computation.  The
result provides a very accurate description of the $dt$ fusion
reaction that involves only three real parameters. Our simple
description may be employed in  calculations of
plasma screening effects that employ field theory methods and thus
requires a field theory of the fusion process \cite{screen}.
(This was the initial motivation for our work on this theory.) 

\subsubsection{Results}
\begin{figure*}[t]
\vskip 1cm
\includegraphics[scale=1.4]{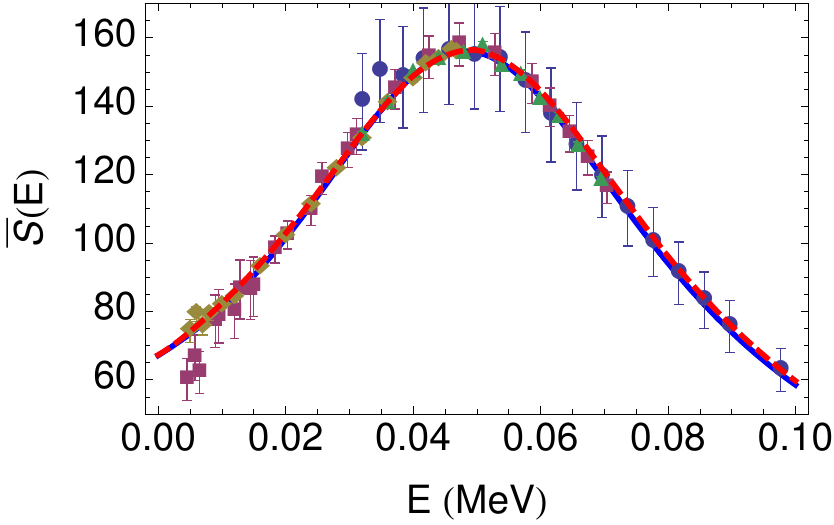}
\null
%\vskip 0.5cm 
\caption{%\captionskip
Dimensionless version of the astrophysical factor 
$\overline S_{dt\to n\alpha}$ determined by the definition 
(\ref{astrodef}) for the $dt$ reaction 
compared with the experimental data as a function of the deuteron center-of-mass
energy $E$. The solid (blue) curve is the best fit of the simple 
effective field theory result (\ref{ourresult}). It has a $\chi^2$
per degree of freedom of 0.784. The 
dashed (red) curve is based on the cross section of Bosch and Hale
\cite{B&H92} . The multilevel, multichannel $R$-matrix analysis of the 
$^5{\rm He}$ system on which the Bosch and Hale cross sections are
based includes data for $n\alpha$ and $dt$ elastic scattering,
in addition to those for the associated inelastic reactions, at
energies equivalent to a laboratory deuteron energy up to 11 MeV. 
It fits the 2665 experimental data points included using 117 free
parameters with a $\chi^2$ per degree of freedom of 1.56. The (magenta)
squares are the data of Arnold {\em et al.} \cite{Arn54}; the (olive) diamonds
are the data of Jarmie {\em et al.} \cite{Jar84} renormalized by a factor of
1.017; the (green) triangles are the relative data of Brown {\em et al.} 
\cite{Brn87}  renormalized by a factor of 1.025. The necessity of these 
renormalizations of the experimental data is discussed in the text.
The (blue) circles are the older data of Argo {\em et al.} \cite{Arg52}
which we show for completeness but which we do not use in our fit. 
}
\label{fig:1}
\end{figure*}
Since the paper contains lengthy calculations, it is
worthwhile to present the $dt$ fusion result here before plunging into all 
the details, including Coulomb corrections.  A major effect of these is
provided by the familiar Gamow barrier penetration factor for the initial
charged deuteron and triton particles.  It is the 
square of the Coulomb wave function evaluated a zero particle separation, 
$\psi^{(C)}_{{\bf p}_{dt}}(0)$,
\begin{equation}
\left| \psi^{(C)}_{{\bf p}_{dt}}(0) \right|^2 = 
\frac{2\pi\eta \, }{\exp\{2\pi\eta\} -1 } \,,
\label{psi0}
\end{equation}
in which for the deuteron and triton, each with a single electron
charge $e$, in ordinary cgs units (but with the $\hbar = 1$ convention
that we usually follow),
\begin{equation}
\eta = \frac{e^2}{v_{dt}} \,,
\label{eeta}
\end{equation}
where $v_{dt}$ is the relative velocity of the deuteron and triton.  We use 
$m_{ab}$ to denote the reduced mass of a pair of particles $a ,b$. So the 
$dt$ momentum in the center-of-mass system is $p_{dt} = m_{dt} \, v_{dt}$, 
with the energy of this relative motion in the center-of-mass system given by
\begin{equation}
E = \frac{1}{2} m_{dt} \, v_{dt}^2 = \frac{{\bf p}_{dt}^2}{2m_{dt}} \,.
\end{equation}
With our convention for the zero point of the energy $W$ in the center-of-mass
system, 
\begin{equation}
\frac{{\bf p}_{dt}^2}{2 m_{dt}}
= W + \epsilon_d + \epsilon_t \,,
\end{equation}
in which $\epsilon_d$ and $\epsilon_t$ are the deuteron and triton binding 
energies. Thus, at a $dt$ threshold where the $dt$ relative momentum vanishes,
$W = - \epsilon_d - \epsilon_t < 0$.  
In our approximation in which the particles interact only with an unstable
intermediate field, in addition to this `external propagation barrier 
penetration', the only other Coulomb corrections are to  
the $dt$ ``bubble graphs'' that appear in the unstable $^5{\rm He}^*$  
propagator.  The inclusion of all Coulomb effects is detailed in the work 
leading to Eq.~(\ref{fusionC}), which reads:
\begin{eqnarray}
\sigma_{dt \to n\alpha}  &=&
\frac{8}{9}  4\pi 
 m_{n\alpha}  \frac{p_{n\alpha}^5}{v_{dt}} 
\frac{g^2_{dt}}{4\pi} \, \frac{g^2_{n\alpha}}{4\pi}   
\left| \psi^{(C)}_{{\bf p}_{dt}}(0) \right|^2 
\left|G_*^{(C)}(W)\right|^2 .
\nonumber\\
&&
\end{eqnarray}
Here $p_{n\alpha}$ is the relative momentum in the center-of-mass frame of the
produced $n , \alpha$ particles.  By energy conservation, it is given by
\begin{equation}
\frac{p_{n\alpha}^2}{2 m_{n\alpha}} = \frac{p_{dt}^2}{2m_{dt}} + Q \,,
\end{equation}
were $ Q \simeq 17.59$ MeV is the energy release of the reaction.  Since the
$n\alpha$ pair is produced in a $D$ wave, the amplitude contains a factor of 
$p_{n\alpha}^2$ and the squared amplitude $p_{n\alpha}^4$.  Phase space of the 
produced particles gives an additional factor of $p_{n\alpha}$, so that an
overall factor of $p_{n\alpha}^5$ appears. The couplings of the initial 
$d , t$ fields and the final $n ,\alpha$ fields to the unstable 
$^5{\rm He}^*$ field are denoted by $g_{dt}$ and $g_{n\alpha}$.

The unstable interacting Green's function that appears in the fusion cross 
section is given by 
\begin{align}
\label{theendd}
&\left| G_*^{(C)}(W)\right|^{-2}  =
\Bigg[ \frac{p_{dt}^2}{2m_{dt}} - E_* 
    - \frac{g_{dt}^2}{4\pi} \, \Delta(W) 
\Bigg]^2
\nonumber\\
&+
\Bigg[ \frac{g_{dt}^2}{4\pi}  \, 2 \, m_{dt} \,
   p_{dt} \, \left| \psi^{(C)}_{p_{dt}}(0) \right|^2
+ \frac{g_{n\alpha}^2}{4\pi} \, \frac{2}{3} \,
      m_{n\alpha} \, p_{n\alpha}^5 \Bigg]^2.
\end{align}
This is the function  given in Eq.~(\ref{theend}) whose derivation and 
description precedes Eq.~(\ref{theend}). The energy $E_*$ along with 
the coupling parameters $g_{dt}^2$ and $g_{n\alpha}^2$ are determined 
by fitting the $dt \to n\alpha$ fusion cross section. 

The function $\Delta$ is a Coulomb-modified $dt$ loop function
that is given by [see Eq.~(\ref{delta})]
\begin{eqnarray}
\Delta(W) &=& 
 \frac{4m_{dt}}{b_0} \,
 \left[ {\rm Re} \, \psi(i\eta) - \ln\eta \right] \,,  
\end{eqnarray}
where $b_0 = 1 / e^2 \, m_{dt} = 24.04 \, {\rm fm}$ and 
$\psi(z) = \Gamma'(z) / \Gamma(z)$ is the logarithmic derivative
of the Gamma function. Although the peak in the fusion cross section,
or the maximum of the modified astrophysical factor 
$\overline S_{dt \to n\alpha}$ shown in Fig.~\ref{fig:1}, are determined by
$E_*$, the positions of these are not directly related to the value of $E_*$
since there are shifts brought about by the Coulomb self-energy correction
$\Delta(W)$ and by the variation of the factors that involve
$p_{dt}$ and $p_{n\alpha}$.

The astrophysical `$S$ factor' is conventionally defined by
\begin{equation}
S = E \, e^{2\pi\eta} \, \sigma  \,.
\end{equation}
The multiplication by $ E = m v^2 /2 $ removes the two factors of 
$ 1 / v $ that naturally appear in the cross section: the $ 1 / v $
arising from the division of the reaction rate by the incident flux, 
and the $1 / v $ factor that appears in the overall $\eta$ factor in
the squared Coulomb wave function (\ref{psi0}).  The factor 
$\exp\{ 2 \pi \eta \}$ removes the major exponential factor (the factor 
which appears in the Gamow barrier penetration approximation) in the 
squared Coulomb wave function (\ref{psi0}). We prefer, however,  to 
use a slightly modified astrophysical `$\overline S$ factor' that 
we define by
\begin{eqnarray}
\overline S_{dt \to n\alpha} &=& 
  \frac{p_{dt}^2}{\hbar^2 } \, \left[ e^{2\pi\eta} - 1 \right] \, 
     \sigma_{dt \to n\alpha}  
\nonumber\\
            &=&
\frac{2 \, m_{dt}}{\hbar^2} E \, 
        \left[ e^{2\pi\eta} - 1 \right] \, \sigma_{dt \to n\alpha} \,.  
\label{astrodef}
\end{eqnarray}
Here we have multiplied by $p^2$ rather than by $E = p^2 / 2m $ because
this makes $\overline S$ dimensionless\footnote{We have displayed the
$\hbar$ factors explicitly so as to emphasize that we are multiplying
by a wave number squared, $(p/\hbar)^2 \sim ({\rm Length})^{-2}$, although
in general we use quantum units in which $\hbar = 1$.}.  Moreover, we have
multiplied by $[\exp\{2\pi\eta\} - 1]$ rather than by only the Gamow
barrier penetration factor $\exp\{2\pi\eta\}$ so as to remove the complete
energy dependence of the squared Coulomb wave function\footnote{We are
interested in the energy range $0 < E < 300 $ keV with includes the
resonance at $E \simeq 50$ keV. The change from $S$ to $\overline S$ is 
of relative order $\exp\{-2\pi\eta\}$ and increases as the energy
increases.  At $E = 300$ keV, the change is about 10\%.}.  

In terms of this notation, our result becomes
\begin{align}
\label{ourresult}
\overline S_{dt \to n\alpha}  &=
\frac{8}{9} \, 4\pi \, m_{dt} \, m_{n\alpha} \,   p_{n\alpha}^5 \,\, 
\frac{g^2_{dt}}{4\pi} \, \frac{g^2_{n\alpha}}{4\pi}  \,\, 
\frac{2\pi}{b_0}  \,\, \nonumber \\
&\times \left|G_*^{(C)}(W)\right|^2 \,.
\end{align}
A fit of this result to the data reduced to construct 
$\overline S_{dt \to n\alpha} $ is presented in Fig.~\ref{fig:1}.
The fit to the $dt$ fusion cross section
with our formula gives the parameter values 
\begin{align}
   E_* &= - 154 \, \pm 8 \,\, {\rm keV} \,, \nonumber \\
\frac{g_{dt}^2}{4\pi} &= 199 \, \pm 8 \,\, {\rm fm}^3 \, {\rm MeV}^2 \,, 
   \nonumber \\
   \label{eqn:pars}
\frac{g_{n\alpha}^2}{4\pi} &= 16.4 \, \pm 1.0 \,\, {\rm fm}^7 \, 
{\rm MeV}^2 \,.
\end{align}

The early cross-section measurements \cite{Arn54,Arg52} used to
determine the parameters of the fit were reported with rather large
uncertainties (typically $\sim 10$\%), which combined relative
and normalization (scale) uncertainties. However, in the more recent
measurement of Jarmie {\em et al.} \cite{Jar84}, the relative errors
were much smaller ($\sim 0.5$\%), and were reported separately from
the larger scale uncertainty of 1.26\%. The subsequent measurement of Brown
{\em et al.} \cite{Brn87} likewise had small relative errors, but no
absolute normalization was determined in this experiment.  For the
purpose of reporting the data, Brown and {\em et al.} determined an
approximate scale by matching in the region of overlap to the earlier
absolute measurement of Jarmie {\em et al.}.

When fitting these data in the comprehensive $^5$He $R$-matrix analysis that
was used to produce the reaction cross sections of Bosch and Hale
\cite{B&H92}, separate normalization parameters were allowed to vary
for each data set, the one for the Jarmie data being constrained in
the total $\chi^2$ by its 1.26\% uncertainty, and the one for the
Brown {\em et al.} data unconstrained, since it was purely a relative
measurement.  The values of the renormalization factors found from
that analysis, 1.017 for Jarmie {\sl et al.}  \cite{Jar84}, and 1.025
for Brown {\em et al.} \cite{Brn87}, were applied to the experimental
data sets (cross sections and uncertainties) prior to performing the
more limited fitting of effective field theory result  
$\bar{S}_{dt\rightarrow n\alpha}$ [Eq.~(\ref{ourresult})] over the
resonance described here\footnote{We also tried letting the normalizations
on these data sets float in this latter analysis, and they varied from the
values given above by about 0.25\%, well within the expected variance
of these numbers, so that there was no need to employ this different scale.}.

Our result, which entails only three parameters, fits the data 
very well.  To achieve this, it is necessary to start with a 
free-particle Lagrangian for the unstable $^5{\rm He}^*$ field with the 
``wrong'' sign. This would not be acceptable if the theory were taken to 
be more fundamental with an extended region of validity rather than 
a effective theory whose applicability is only to the low-energy regime. 
It is easy to show that the simple theory with two initial spin zero
particles which interact via an intermediate (``s-channel'') field 
(the simple scalar-particle analog of our theory) produces an effective 
range formula with a negative effective range parameter \cite{BHP}. A positive
effective range parameter is achieved in this theory if the intermediate
field has a wrong-sign free-particle Lagrangian. Thus the restricted 
validity of this simple effective field theory should be acceptable  
just as is that of the effective range theory\footnote{Kaplan 
\cite{DK} has obtained a good fit to the neutron-proton singlet $S$-wave 
scattering phase shift out to a lab energy of 340 MeV with a theory that
contains pion exchange, a local point (contact) interaction, and an
s-channel intermediate field with a wrong-sign free-particle Lagrangian.
Schwinger \cite{Julian} described the deuteron as an effective field and
derived the effective range formula for the neutron-proton triplet
$S$-wave scattering as well as the corresponding approximation for the
deuteron form factor and the low-energy deuteron photodisintegration. 
A careful reading of \cite{Julian} reveals that,
hidden in the non-relativistic reduction of a relativistic theory that
involves a wave function renormalization for the deuteron field, the
resulting free-particle deuteron propagator corresponds to a wrong-sign
Lagrangian, a sign change brought about by a negative sign in the
wave function renormalization.}.

\subsubsection{Outline}
\label{sssec:outline}

Section \ref{kinship} explains our conventions for the fields and their
free-particle Lagrange functions. Section \ref{unstable} defines the 
interaction Lagrange functions for the coupling of the initial $dt$
and the final $n\alpha$ to the unstable $^5{\rm He}^*$ field and
the appropriate spin-orbit combination of the $n\alpha$ fields that enter
into their interaction.

Section \ref{effQFT} describes the dynamics of our theory in the absence 
of the Coulomb interactions. This is done in some detail because this 
underlying theory, which involves higher spin fields, has some complexity,
and it clarifies the development to proceed with simpler stages. 
Section \ref{*prop} presents the calculation of the self-energy functions
for the unstable $^5{\rm He}^*$ propagator using dimensional continuation
to define their needed regularization and express the intermediate 
expressions in term of quantum-mechanical transformation functions that
simplifies the subsequent computation of Coulomb corrections.
Section \ref{reactamp} describes our result for the
$dt \to n\alpha$ fusion cross section in the absence of Coulomb interactions.

Section \ref{CCcorr} displays the Coulomb corrections to the fusion 
process.  The $dt$ particles initially interact at a point, thereby
bringing about a factor of the squared Coulomb wave function at the
origin $|\psi^{(C)}(0)|^2 $.  The $dt$ piece of the resonant state
propagator also has Coulomb corrections. Using the formalism developed
in Sec.~\ref{*prop}, these corrections become expressed in a dispersion
relation form that is a representation of the logarithmic derivative
of the Gamma function $\psi(z)$.  

After a brief summary of our work in the concluding Section
\ref{sec:conc}, unsuccessful approaches to avoid the introduction of
the wrong-sign $^5{\rm He}^*$ free propagator are mentioned. We then note
how extensions of the effective field theory method to multichannel
descriptions of light nuclear reactions might be obtained without
great effort.

Appendix \ref{galinv} contains a short account of Galilean invariance
that provides results needed in the text. Efficient quantum field theory 
methods that couple spins are explained in Appendix \ref{spin}. The
theory of Coulomb corrections is discussed in Appendix \ref{coul}.

\section{Effective Field Theory: Ingredients}

\subsection{Fields, Kinematics}
\label{kinship}

As discussed in the Introduction, each particle in our reaction system
is described by creation and annihilation fields. The free-field part
of the Lagrange function for each of these fields has the generic form
\begin{equation}
{\cal L}^{(0)}_\smA = 
\chi^\dagger_\smA \,  i \, \frac{\partial}{\partial t} \, \chi_\smA 
  - {\cal H}^{(0)}_\smA \,, 
\end{equation}
with
\begin{equation}
{\cal H}^{(0)}_\smA = \chi^\dagger_\smA \left[ \frac{ - \nabla^2}{2m_\smA} 
                - \epsilon_\smA \right] \chi_\smA \,.
\label{H0A}
\end{equation}
As shown in Appendix \ref{galinv}, Galilean invariance requires that
the inertial mass $m_\smA$ of a composite nucleus is the sum of the
\begin{table*}[t]
\begin{tabular*}{0.75\textwidth}{@{\extracolsep{\fill}}lcccc}
   Particle & Spin & Operators & Mass & Binding \\
   \hline
   Alpha & $0^+$ & 
   $\phi^\dagger_\alpha({\bf r},t),\ \phi_\alpha({\bf r},t)$ &
   $m_\alpha = 2 m_p + 2 m_n$ & $\epsilon_\alpha$ \\
   Deuteron & $1^+$ & 
   ${\bm \phi}_d^\dagger({\bf r},t),\ \bm{\phi}_d({\bf r},t)$ 
   & $m_d = m_p + m_n$ & $\epsilon_d$
\end{tabular*}
\caption{\label{tab:bosons}Bosonic fields and their properties.}
\end{table*}
\begin{table*}[ht]
\begin{tabular*}{0.75\textwidth}{@{\extracolsep{\fill}}lcccc}
   Particle & Spin & Operators & Mass & Binding \\
   \hline
    Neutron & $\tfrac{1}{2}^+$ & $\psi_n^\dag({\bf r},t)$,
   $\psi_n({\bf r},t)$ & $m_n$ & $\epsilon_n\equiv 0$ \\
    Triton & $\tfrac{1}{2}^+$ & $\psi_t^\dag({\bf r},t)$,
   $\psi_t({\bf r},t)$ & $m_t=m_p+2m_n$ & $\epsilon_t$ \\
    $^5$He$^*$ & $\tfrac{3}{2}^+$ & $\psi_*^\dag({\bf r},t)$,
   $\psi_*({\bf r},t)$ & $m_*=2m_p+3m_n$ & $\epsilon_*$
\end{tabular*}
\caption{\label{tab:fermions}Fermionic fields and their properties.}
\end{table*}
masses of the neutrons and protons of which it is composed.  We
have written $\epsilon_\smA > 0 $ for the binding energy of the 
composite particle $\smA$. The total free particle Hamiltonian,
\begin{equation}
H^{(0)} = {\sum}_\smA \, \int(d^3{\bf r}) \, {\cal H}^{(0)}_\smA \,,
\label{totH0}
\end{equation}
measures the energy of an asymptotic state where all the particles
a separated by large distances.

In view of the structure of the total Hamiltonian (\ref{totH0}) with 
the pieces (\ref{H0A}), the total energy $E_{ab}$ of a pair $a\,,\,b$ 
of stable particles separated by large distances is
\begin{equation}
E_{ab} = \frac{{\bf p}_a^2}{2m_a} - \epsilon_a
         + \frac{{\bf p}_b^2}{2m_b} - \epsilon_b =
\frac{{\bf P}_{ba}^2}{2M_{ab}} + \frac{{\bf p}_{ba}^2}{2m_{ab}}
- \epsilon_a - \epsilon_b \,,
\end{equation}
in which $M_{ab}$ and $m_{ab}$ are the total and reduced masses of the
$a\,,\,b$ system; ${\bf P}_{ba}$ and ${\bf p}_{ba}$ are the total and
relative momenta. 
The energy $W_{ab}$ in the center-of-mass system is the Galilean
invariant
\begin{equation}
W_{ab} =  E_{ab} - \frac{{\bf P}_{ba}^2}{2M_{ab}} 
= \frac{{\bf p}_{ba}^2}{2m_{ab}} - \epsilon_a - \epsilon_b \,.
\end{equation}

Our description of the $ d \, t \to n \, \alpha$ reaction will employ
only the initial and final particles, the deuteron ($d$), neutron
($n$), and the triton ($t$) alpha ($\alpha$), and a single unstable
$^5$He$^*$ nucleus. The corresponding bosonic fields and their
properties: spin-parity, masses, and binding energies, are listed in
Table \ref{tab:bosons}.  Note that we conveniently describe the
deuteron spin states as a vector representing ``linear polarization'';
the usual ${J_z}' = m = \{\pm 1,0\}$ states are the $\{(x \pm i
y)/\sqrt2, z \}$, components of this vector. Properties of the
fermionic fields are given in Table \ref{tab:fermions}.

As explained in Appendix~\ref{spin}, the condition that the unstable
$^5{\rm He}^*$ field carry only spin $3/2$ (with no additional spin
$1/2$ piece) can be conveyed in the requirement that this vector-spinor
field obeys\footnote{This is just the non-relativistic version of the
Rarita-Schwinger description of spin $3/2$ fields \cite{RS}.} 
\begin{equation}
\bm{\sigma} \cdot \bm{\psi}_*({\bf r},t) = 0 
= \bm{\psi}_*^\dagger({\bf r},t) \cdot \bm{\sigma} \,,
\end{equation}
in which $\bm{\sigma}$ are the Pauli spin matrices. 

The unstable, resonant $^5{\rm He}^*$ state, has a `binding' energy 
$ \epsilon_*$ that is negative so that can decay into a deuteron plus 
a triton.  The conservation of total energy $W$ in the center-of-mass 
system for the $dt$ fusion reaction gives
\begin{equation} 
\frac{{\bf p}_{dt}^2}{2m_{dt}} - \epsilon_d - \epsilon_t =
\frac{{\bf p}_{n\alpha}^2}{2m_{n\alpha}} - 
     \epsilon_n - \epsilon_\alpha \,.
\end{equation}
At threshold, $ {\bf p}_{dt} = 0 $, and the produced $n , \alpha$
pair has a kinetic energy 
$ {\bf p}^2_{n\alpha}/2m_{n\alpha} = Q $.  Here $Q$ is the conventional
notation for the energy liberated by the reaction. 
Since by our convention $\epsilon_n = 0$, 
\begin{equation}
Q = \epsilon_\alpha - \epsilon_d - \epsilon_t \simeq 17.59 
\,\,  {\rm  MeV} .
\end{equation}

\subsection{Unstable Particle Interactions}
\label{unstable}

As discussed in the Introduction, the interaction Lagrange function
describes the coupling of the reacting particles to an intermediate
unstable field that describes a $3/2^+$ resonance 
$^5$He$^*$ in the intermediate state: 
\begin{align} 
\label{int}
{\cal L}_1 &=
g_{d t} \left[ \bm{\psi}_*^\dagger \,\, \psi_t \cdot \bm{\phi}_d
+ {\bm{\phi}_d}^{\, \dagger} \cdot \psi^\dagger_t \, \bm{\psi}_* \right] 
\nonumber \\
&+ g_{n\alpha} \left[ \bm{\psi}_*^\dagger \cdot \bm{\Psi}_{\alpha n}
        + \bm{\Psi}^\dagger_{\alpha n} \cdot \bm{\psi}_* \right]
\end{align}
Here the $dt$ field pair contains spin $1/2$ as well as spin $3/2$. 
However, the coupling of this pair to the unstable particle field 
with spin $3/2$ projects out only the spin $3/2$ part of the $dt$
pair. The coupling of the unstable particle field to the $\alpha n$
pair is more complicated since as discussed in detail in 
Appendix~\ref{spin}, it involves an internal D-wave angular momentum 
in this pair. This $l=2$ internal angular momentum combines with
the spin $1/2$ in the neutron to produce the spin ${3/2}^+$ field
$\bm{\Psi}_{\alpha \, n}$.  As explained in Appendix~\ref{spin}, this field
is given by
\begin{equation}
\Psi^l_{\alpha \, n}({\bf r},t) = 
\phi_\alpha({\bf r},t) \, {\cal T}^{lm}_{\alpha n} \,
\sigma^m \, \psi_n({\bf r},t) \,,
\label{ugh}
\end{equation} 
in which a sum over repeated vector or tensor indices is implied, and with
\begin{equation}
{\cal T}^{lm}_{\alpha n} = {\cal P}^l_{\alpha n} {\cal P}^m_{\alpha n}
-\frac{1}{3} \, \delta^{lm} \, {\cal P}^k_{\alpha n} {\cal P}^k_{\alpha n} \,,
\label{calT}
\end{equation}
where
\begin{equation}
{\cal P}^k_{\alpha n} =
\frac{m_{n\alpha }}{m_n} \frac{1}{i}\stackrel{\rightarrow}{\nabla}^k -
\frac{m_{n\alpha}}{m_\alpha} \frac{1}{i} \stackrel{\leftarrow}{\nabla}^k \,,
\end{equation}
with $m_{n\alpha }$ the reduced mass of the alpha-neutron system. The arrow
over a derivative indicates whether the derivative acts to the left or to 
the right. As we shall
see, the differential operator ${\cal P}^k_{\alpha n}$ reduces to the relative
momenta of the $n\alpha$ pair when the reaction amplitudes are computed.

\section{Effective Field Theory: Dynamics}
\label{effQFT}

The reaction amplitudes may be described by the interaction picture
which involves free-field matrix elements of the time-ordered,
unitary evolution operator
\begin{equation}
U = \left( \exp\left\{ i \int dt \int (d^3{\bf r}) \, 
{\cal L}_1 \right\} \right)_+ \,.
\label{U}
\end{equation}
The propagators of non-relativistic fields are retarded functions in
time: particles are created at an earlier time and then destroyed at a
later time. Hence, on expanding the interaction picture time evolution
operator (\ref{U}), it is easy to see that the only contributions to a
two-body reaction with an interaction of the form of Eq.~(\ref{int})
are as follows. Starting at an early time, the initial particle pair
is destroyed and the unstable $^5$He$^*$ particle is created.
In the leading order expansion of the evolution operator, this
unstable particle decays into the particles in the final state.  In
the next-to-leading order, the unstable particle propagates for some
time and then decays into a particle pair.  Each particle in this pair
now propagates for another period of time until the pair is destroyed
with the creation of another unstable particle.  This chain of
``bubbles'' goes on until the final two-particle state is reached.
The first terms in the expansion of the evolution operator give a
single ``bubble'' surrounded by two unstable free-field propagators.
The next set of expansion terms give two ``bubbles'' joined by three
unstable free-field propagators. And so forth to result in an infinite
set of graphs consisting of alternating lines and ``bubbles''. 

We consider in the next subsection, Sec.~\ref{*prop}, the unstable
particle Green's function neglecting the Coulomb interaction. This
allows us to focus on the evaluation of the $n\alpha$ and $dt$
contributions to the self-energy, $\Sigma_{n\alpha}(W)$ and
$\Sigma_{dt}(W)$. These are sufficiently complex computations, even
with the neglect of the instantaneous Coulomb interactions, to warrant
a dedicated discussion. Then, in following Sec.~\ref{CCcorr}, we
include corrections due to the instantaneous Coulomb interactions for
these particles.

\begin{figure*}
   \includegraphics[height=1.6cm]{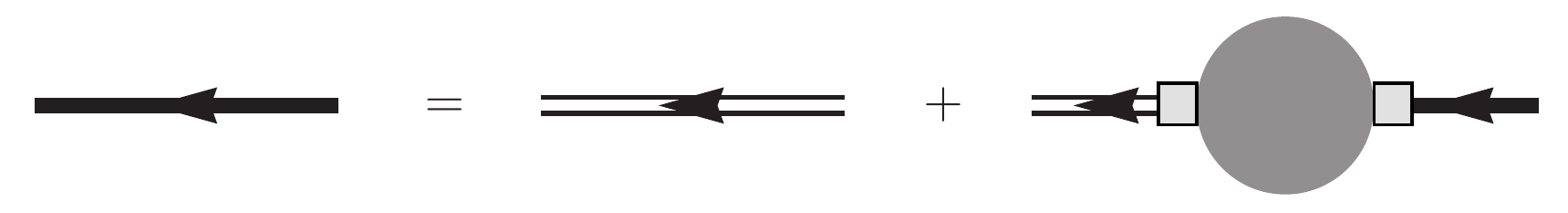}
   \caption{\label{fig:5heprop} 
      Diagrammatic structure of the interacting $^5$He$^*$ Green's
      function. The thick, directed line represents the interacting 
      Green's function with all its self-energy corrections. The double 
      line stands for the wrong-sign free-particle propagator. The shaded 
      region immediately to the right of the free propagator represents 
      the $n\alpha$ and $dt$ self-energies contained in the $\Sigma(W)$
      that appears in Eq.~\eqref{eqn:sig}. 
      }
   \includegraphics[height=3cm]{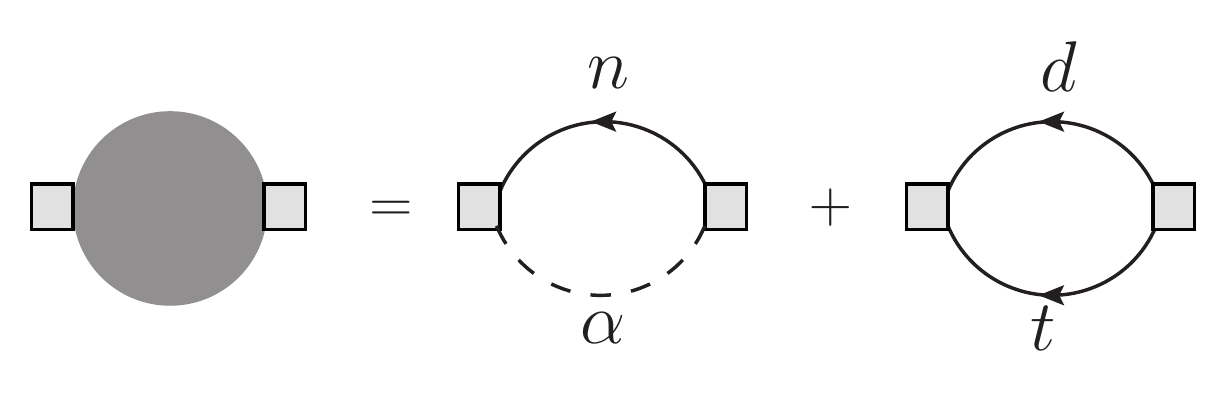}
   \caption{\label{fig:dtse-nocoul}The self-energy diagrams of the 
      $^5$He$^*$, neglecting the instantaneous Coulomb interaction, 
      corresponding to Eq.~\eqref{eqn:sig}. Shaded boxes indicate the 
      $g_{n\alpha}$ and $g_{dt}$ vertices appropriate to each graph.  
      The first loop graph on the right-hand side describes the 
      $n\alpha$ contribution Eq.~\eqref{eqn:sna}; the second loop graph 
      corresponds to the $dt$ contribution Eq.~\eqref{eqn:sdt}.}
\end{figure*}

\subsection{Unstable Particle Green's Function}
\label{*prop}

The unstable particle's Green's function may be expressed as
\begin{align}
\label{unstableprop}
G_*^{lm}({\bf r} - {\bf r}', t-t') &= 
\int \frac{(d^3{\bf p})}{(2\pi)^3} \frac{dE}{2\pi} \,
e^{ i {\bf p} \cdot ( {\bf r} -  {\bf r}' )
- i E (t - t') }
\nonumber \\ &\times 
P_{3/2}^{lm} \, G_*(W) \,.
\end{align}
Here $P_{3/2}^{lm}$ is the projection matrix (\ref{P3/2}) into the 
spin $3/2$ subspace. It is a matrix in the $2\times2$ spinor space 
and a second-rank tensor in the vector indices exhibited. Because of
Galilean invariance, the scalar factor $G_*(W)$ is a function only of
the energy in the center-of-mass frame
\begin{equation}
W = E - \frac{p^2}{2m_*} \,.
\end{equation}

The unstable $^5{\rm He}^*$ inverse Green's function scalar factor has 
the form
\begin{equation}
G_*^{-1}(W) = -(W + \epsilon_*) - \Sigma(W+i\eta) \,,
\label{Gstar}
\end{equation}
where, as discussed at some length in the Introduction and Summary
above, the free-particle piece is taken corresponding to a wrong-sign
Lagrangian. The structure of the corresponding Green's function is 
described by an integral equation in space-time which reduces to an 
algebraic equation in momentum-frequency space as indicated by the 
diagram in Fig.~\ref{fig:5heprop}.  The self-energy $\Sigma(W)$ is 
the sum of a $dt$ part and an $n\alpha$ part:
\begin{equation}
\label{eqn:sig}
\Sigma(W) = \Sigma_{dt}(W) + \Sigma_{n\alpha}(W) \,.
\end{equation}
The self-energy functions, corresponding to $dt$ and $n\alpha$ loop
graphs, is displayed in Fig.~\ref{fig:dtse-nocoul}.

The $dt$ contribution involves the propagator or Green's function for 
the spin-one deuteron which has the simple tensor structure 
$ G_d^{kl} = \delta^{kl} \, G_d $, where $G_d$ is a scalar function, 
and the Green's function for the triton which is a unit matrix in the
$2\times2$ spinor space times a scalar function $G_t$. The unit 
tensor $\delta^{kl}$ describing the deuteron spin and the unit matrix
in the deuteron spin space both act as unity when acting upon the 
components of the unstable field Green's function.  Hence, the $dt$
loop can be written as the scalar function
\begin{equation}
\Sigma_{dt}(W) = 
i \,  g_{dt}^2 \, \int (d^3\bar{\bf r}) \, dt \,
 e^{-i {\bf p} \cdot \bar{\bf r} + i E \, t}
\, G_d(\bar{\bf r},t) \, G_t(\bar{\bf r},t) \,.
\end{equation}
The scalar part of the Green's functions have the generic form 
\begin{align}
&G({\bf r} - {\bf r}', t-t') =
 -i \Big\langle 0 \Big| \left( \chi({\bf r},t)  \chi^\dagger({\bf r}',t') 
         \right)_+ \Big| 0 \Big\rangle
\nonumber\\
&= -i \theta(t-t') \int \frac{(d^3{\bf p})}{(2\pi)^3} 
     e^{i {\bf p} \cdot ( {\bf r} - {\bf r}' ) } 
       e^{i E(p)  ( t - t') } ,
\end{align}
in which $\theta$ is the unit step function and 
\begin{equation}
 E(p) = \frac{{\bf p}^2}{2m} - \epsilon 
\label{nucE}
\end{equation}
is the energy which includes the binding energy $-\epsilon$ as well as
the kinetic energy with a mass $m$ that is the sum of the nucleon masses
that make up the nucleus described by the field $\chi$.

Such Green's functions have the structure of a time-dependent,
quantum-mechanical transformation function of a free particle. It will 
prove convenient to write them in this form,  namely as
\begin{equation}
G({\bf r} - {\bf r}', t-t') = 
-i \langle {\bf r} , t | {\bf r}' , t' \rangle^{(0)} 
\, \theta(t-t') \,.
\end{equation}
Here we have included the superscript to indicate that we are
evaluating the free-particle transformation function with no 
Coulomb interactions. Later in Sec.~\ref{CCcorr}, when we
turn to the Coulomb corrections, this superscript will indicate the
evaluation of the transformation function in the presence of the
Coulomb interaction with $(0)\to(C)$.  It is useful to use this
relation because it is then natural to pass to center-of-mass and
relative coordinates and write
\begin{align}
   &G_b({\bf r}_b - {\bf r}_b' , t - t') \, 
G_a({\bf r}_a - {\bf r}_a' , t - t')\nonumber \\
&= 
- \langle {\bf R} , t | {\bf R}' , t'  \rangle^{(0)}_{ ba \,\,\rm CM} \,\,
\langle {\bf r} , t | {\bf r}' , t' \rangle^{(0)}_{ba \,\, \rm rel} 
\,\, \theta(t-t') \,.
\label{Gab}
\end{align}
Here, as usual,
\begin{equation}
{\bf R} = \frac{m_a {\bf r}_a + m_b {\bf r}_b }{m_a + m_b} \,,
\qquad\qquad {\bf r} = {\bf r}_b - {\bf r}_a 
\end{equation}
are the center-of-mass and relative coordinates. 
The free-particle dynamics in the transformation function of the
relative motion 
$\langle {\bf r} , t | {\bf r}' , 0 \rangle^{(0)}_{ba \,\, \rm rel} $
is described by the Hamiltonian
\begin{equation}
H_{ba \,\, \rm rel} = \frac{{\bf p}_{ba}^2}{2 m_{ab}} 
- \epsilon_b - \epsilon_a 
\label{free2}
\end{equation}
that contains the binding energies displayed in Eq.~(\ref{nucE}) so as 
to provide the correct reference energy. 
We shall find this decomposition helpful when we compute the Coulomb 
corrections to the reactions\footnote{With Coulomb interaction present
in the initial $dt$ channel, the initial four-point Green's function
does not factor into the product of 2 two-point functions. This is
discussed in detail in Appendix \ref{coul}.}. 
To return to the evaluation of
the  self-energy function $\Sigma_{dt}$, we note that at the coincident
points ${\bf r}_d = {\bf r}_t = \bar{\bf r}$, ${\bf R} = \bar{\bf r}$,
and we encounter
\begin{align}
   &\int (d^3\bar{\bf r}) \, e^{- i {\bf p} \cdot \bar{\bf r} }
\langle \bar{\bf r} , t | {\bf 0} ,0 \rangle^{(0)}_{ ba \,\,\rm CM} 
\nonumber \\
&= \exp\left\{ -i \frac{{\bf p}^2}{2 M_{ba}} \, t \right\} \,.
\end{align}
In the present case, $M_{ba} = m_d + m_t = 2 m_p + 3 m_n = m_*$.
Hence the self-energy function involves a Fourier transform in time
with a single energy variable $W = E - {\bf p}^2 / 2 m_*$, as must be
the case in virtue of the Galilean invariance of the theory, and we have 
\begin{align}
\Sigma_{dt}(W) 
&=
-i g_{dt}^2 \int_0^\infty dt \, e^{ i W t }
\langle {\bf 0} , t | {\bf 0} , 0 \rangle^{(0)}_{dt \,\, \rm rel} \,.
\label{dtselfE}
\end{align}

To evaluate the loop function that appears in the 
self-energy $\Sigma_{n\alpha}(w)$, we note that it entails 
\begin{align}
   &\left\langle 0 \left| 
\Psi^k_{\alpha \, n}(\bar{\bf r},t)
\Psi^l_{\alpha \, n}(\bar{\bf r}',t)  
\right| 0 \right\rangle 
\nonumber\\
&=
\left\langle 0 \left| 
\phi_\alpha(\bar{\bf r},t) \, {\cal T}^{km}_{\alpha n}
\sigma^m \psi_n(\bar{\bf r},t) 
 \psi_n^\dagger(\bar{\bf r}',t) \sigma^n
 {\cal T}^{nl}_{\alpha n} \phi_\alpha^\dagger(\bar{\bf r}',t) 
\right| 0 \right\rangle
\nonumber\\
&=
\langle \bar{\bf r} , t | \bar{\bf r}' , t'  \rangle^{(0)}_{ ba \rm CM} 
\sigma^m \sigma^n
\left.
\left[ \nabla^k \nabla^m - \frac{1}{3} \delta^{km} \nabla^2 \right]
\right.
\nonumber\\ &\times
\left.
\left[ \nabla^n \nabla^l - \frac{1}{3} \delta^{nl} \nabla^2 \right]
\langle {\bf r} , t | {\bf 0} , 0 \rangle^{(0)}_{ba \rm rel} 
\right|_{{\bf r} = {\bf 0} }.
\label{nalpha}
\end{align}
Here we have made use of the translational invariance of the 
free-particle transformation function
$ \langle {\bf r} , t | {\bf r}' , 0 \rangle^{(0)}_{ba \,\, \rm rel} 
=
\langle {\bf r} - {\bf r}', t | {\bf 0} , 0 \rangle^{(0)}_{ba \,\, \rm rel} 
$
to write all the derivatives on the left as shown.  Since the 
${\bf r} \to {\bf 0}$ limit yields a rotationally invariant function
whose tensor structure can only involve the unit tensor $\delta^{^{{\bf ..}}}$,
we have
\begin{align} 
   &\left[ \nabla^k \nabla^m - \frac{1}{3} \delta^{km} \nabla^2 \right]
   \nonumber \\
   &\times\left[ \nabla^n \nabla^l - \frac{1}{3} \delta^{nl} \nabla^2 \right]
\langle {\bf r} , t | {\bf 0} , 0 \rangle^{(0)}_{ba \rm rel} 
\Big|_{{\bf r} = {\bf 0} }.
\nonumber\\
&= \frac{1}{15} \left[ \delta^{kn} \delta^{ml} + 
 \delta^{kl} \delta^{mn}  -\frac{2}{3} \delta^{km} \delta^{nl} \right]
 \nonumber \\ &\times
\left( \nabla^2 \right)^2
\langle {\bf r} , t | {\bf 0} , 0 \rangle^{(0)}_{ba \rm rel} 
\Big|_{{\bf r} = {\bf 0} },
\end{align}
as contractions with various $\delta^{^{{\bf ..}}}$ establishes. Placing 
this result in Eq.~(\ref{nalpha}), using the Pauli matrix composition law 
(\ref{Pauli}) and the form (\ref{P3/2}) of the spin $3/2$ projection 
matrix, we conclude that
\begin{align}
   &\left\langle 0 \left| 
\Psi^k_{\alpha \, n}(\bar{\bf r},t)  \, \Psi^l_{\alpha \, n}({\bf 0},0)  
\right| 0 \right\rangle \nonumber \\
&=
\langle \bar{\bf r} , t | {\bf 0} , 0  \rangle^{(0)}_{ \alpha n \,\,\rm CM}
 \,\, \frac{1}{3} \, P^{kl}_{3/2} \,
 \nonumber \\
 &\times \left( \nabla^2 \right)^2 \,
\langle {\bf r} , t | {\bf 0} , 0 \rangle^{(0)}_{\alpha n \,\, \rm rel} 
\Big|_{{\bf r} = {\bf 0} } \,.
\end{align}
The projection matrix $  P^{kl}_{3/2} $ that appears here can be omitted.
It either acts upon the $^5{\rm He}^*$ propagator that only contains
spin $3/2^+$. Hence it can be replaced by unity, and 
just as in the previous evaluation of the $dt$ self-energy function, we
have
\begin{align}
\Sigma_{n\alpha}(W) 
&=
-i g_{n\alpha}^2 \int_0^\infty dt e^{ i W t } 
\nonumber \\ &\times
\left( \nabla^2 \right)^2 
\langle {\bf r} , t | {\bf 0} , 0 \rangle^{(0)}_{\alpha n \rm rel} 
\Big|_{{\bf r} = {\bf 0} }.
\label{nalphaselfE}
\end{align}

To complete the computation, we need to evaluate expressions that are
divergent in three spatial dimensions.  These divergences can be
removed by `subtractions' --- by  deleting the divergent pieces
and replacing them by appropriate mass and wave function
renormalizations.  In our case, with the highly divergent $n\alpha$
piece, this subtraction method is a cumbersome method.  Moreover, it
is not a proper, well-defined mathematical procedure.  The proper
procedure is to first regulate the theory to make it well defined, and
then perform whatever renormalizations that are needed.  Any
regularization scheme must make the theory unphysical in some sense
because if it were not, one could have a well-defined, finite theory,
and one should use this new theory rather than that which one started
with. One regularization scheme is that of Pauli and Villars.  It
produces a regularized theory with sectors that have negative
probabilities until the renormalizations are made and the
regularization removed.  Here we shall find it very convenient to use
dimensional regularization where the three spatial
dimensions are continued to an arbitrary $\nu$ dimensional space, and
the limit $\nu \to 3$ is performed only after all computations have
been performed.

In $\nu$ spatial dimensions
\begin{align}
   &\left\{
\begin{array}{c}
1 \\
\left( \nabla^2 \right)^2
\end{array}
\right\}
\left. 
\langle {\bf r} , t | {\bf 0},0 \rangle^{(0)}_{ba \, \rm rel}\right|_{{\bf r} 
= {\bf 0}}
\nonumber \\
&=
e^{i (\epsilon_b + \epsilon_a) \, t} \int \frac{(d^\nu{\bf p})}{(2\pi)^\nu}
\left\{
\begin{array}{c}
1 \\
\left( {\bf p}^2 \right)^2
\end{array}
\right\}
   \exp\left\{ -i t \, \frac{{\bf p}^2}{ 2m_{ba} } \right\}
\nonumber\\
&=
e^{i (\epsilon_b + \epsilon_a) \, t}  \frac{\Omega_{\nu-1}}{(2\pi)^\nu } 
\int_0^\infty p^{\nu-1}  dp 
\left\{
\begin{array}{c}
1 \\
p^4
\end{array}
\right\}
    \exp\left\{\! -i t  \frac{ p^2 }{ 2m_{ba} } \right\} \! ,
\nonumber\\
&&
\label{nunu}
\end{align}
where in the second line we have passed to hyper-spherical coordinates with
$ \Omega_{\nu-1}$ the area of a unit $\nu -1$ sphere embedded in a 
$\nu$-dimensional space.
We change variables, writing explicitly $i = \exp\{\pi i / 2\}$ so as to be 
able to carefully keep track of the phase, to get
\begin{widetext}
\begin{eqnarray}
&& \left\{
\begin{array}{c}
1 \\
\left( \nabla^2 \right)^2
\end{array}
\right\}
\left. 
\langle {\bf r} , t | {\bf 0} , 0 \rangle^{(0)}_{ba \,\, \rm rel}\right|_{{\bf r} = {\bf 0}}
\nonumber\\
&& \qquad\qquad =
e^{i \, (\epsilon_b + \epsilon_a) \, t} \, \, \frac{\Omega_{\nu-1}}{(2\pi)^\nu } \, 
\left( \frac{2 m_{ba}}{t} \right)^{\nu/2} \, e^{-\nu\pi i / 4} \, 
\frac{1}{2} \, \int_0^\infty u^{\nu/2} \, \frac{du}{u} \,
\left\{
\begin{array}{c}
1 \\
- \left( 2m_{ba} /t \right)^2 \, u^2
\end{array}
\right\}
   \, e^{-u}
\nonumber\\
&& \qquad\qquad =
e^{i \, (\epsilon_b + \epsilon_a) \, t} \, \, 
\frac{\Omega_{\nu-1}}{(2\pi)^\nu } \, 
\left( \frac{2 m_{ba}}{t} \right)^{\nu/2} \, e^{-\nu\pi i / 4} \, 
\frac{1}{2} \,
\left\{
\begin{array}{c}
\Gamma\left(\frac{\nu}{2}\right) \\
- \left( 2m_{ba} /t \right)^2 \, \Gamma\left(\frac{\nu}{2} +2 \right)  
\end{array}
\right\} \,.
\end{eqnarray}
We may now complete the evaluation of the one-loop self-energy functions: 
\begin{eqnarray}
\left\{
\begin{array}{c}
\Sigma_{dt}(W)  \\
\Sigma_{n\alpha}(W)
\end{array}
\right\} 
&=&
-i \, \int_0^\infty \, dt \, e^{ i W \, t }  \,
 \left\{
\begin{array}{c}
1 \,  \\
\left(  1 / 3 \right) \, \left( \nabla^2 \right)^2
\end{array}
\right\}
\, \left\{
\begin{array}{c}
g_{dt}^2 \, \left.\langle {\bf r} , t | 
{\bf 0} , 0 \rangle^{(0)}_{dt \,\, \rm rel}\right|_{{\bf r} = {\bf 0}} \\
g_{n\alpha}^2 \, \left.\langle {\bf r} , t | 
{\bf 0} , 0 \rangle^{(0)}_{n\alpha \,\, \rm rel}\right|_{{\bf r} = {\bf 0}}
\end{array}
\right\} \,.
\end{eqnarray}
The $a,b$ center-of-mass channel momentum is defined by
\begin{equation}
\frac{p_{ba}^2}{2 m_{ab}} = W + \epsilon_a + \epsilon_b \,,
\end{equation}
and so we encounter integrals of the form
\begin{eqnarray}
- i \int_0^\infty dt \, e^{i w t}
\left\{
\begin{array}{c}
t^{-\nu/2} \\
t^{-((\nu/2) +2)}
\end{array}
\right\} 
&=& 
e^{-\pi i \nu/4} \, \int_0^\infty dv \, e^{-v}
\left\{
\begin{array}{c}
w^{(\nu/2) - 1} \, v^{-\nu/2} \\
- w^{(\nu/2) + 1} \, v^{-((\nu/2) +2)}
\end{array}
\right\} 
\nonumber\\
&=& 
e^{-\pi i \nu/4} \, 
\left\{
\begin{array}{c}
w^{(\nu/2) - 1} \, \Gamma(1-(\nu/2)) \\
- w^{(\nu/2) + 1} \, \Gamma(-1 - (\nu/2))
\end{array}
\right\} \,,
\end{eqnarray}
in which
\begin{eqnarray}
w =
\left\{
\begin{array}{c}
\frac{p_{dt}^2}{2m_{dt}} \\
\frac{p_{n\alpha}^2}{2m_{n\alpha}}
\end{array}
\right\} 
\end{eqnarray}
for the two cases. Hence
\begin{eqnarray}
\left\{
\begin{array}{c}
\Sigma_{dt}(W)  \\
\Sigma_{n\alpha}(W)
\end{array}
\right\} 
&=&
\frac{\Omega_{\nu-1}}{(2\pi)^\nu } \, \, e^{-\nu\pi i / 2} \, 
\frac{1}{2} \,
\left\{
\begin{array}{c}
 g_{dt}^2 \,  p_{dt}^\nu  \, \frac{2m_{dt}}{p_{dt}^2} \,
\Gamma\left(\frac{\nu}{2}\right) \Gamma\left(1-\frac{\nu}{2} \right) \\
g_{n\alpha}^2 \, \left( 1 / 3 \right) \,
 p_{n\alpha}^{\nu+4} \, \frac{2m_{n\alpha}}{p_{n\alpha}^2} \,
 \Gamma\left(\frac{\nu}{2} + 2 \right)  \Gamma\left(1 - \left(2 +
 \frac{\nu}{2} \right)\right)
\end{array}
\right\}
\nonumber\\
&=&
\frac{\Omega_{\nu-1}}{(2\pi)^\nu } \, \, e^{-\nu\pi i / 2} \, 
\frac{1}{2} \,
\left\{
\begin{array}{c}
 g_{dt}^2 \, p_{dt}^\nu  \, \frac{2m_{dt}}{p_{dt}^2} \,
\frac{\pi}{\sin \pi \left( \frac{\nu}{2} \right)} \\
g_{n\alpha}^2 \, \left( 1 / 3 \right) \,
 p_{n\alpha}^{\nu+4} \, \frac{2m_{n\alpha}} {p_{n\alpha}^2} \,
 \frac{\pi}{\sin \pi\left(\frac{\nu}{2} + 2 \right)} 
\end{array}
\right\} \,.
\end{eqnarray}
\end{widetext}
We now find that there is no impediment to taking the limit $\nu \to 3$.
This is a great advantage of the application of the dimensional 
continuation\footnote{The previous exposition is meant to be self-contained.
If the reader needs a more extended discussion, a full, pedagogical 
description of the dimensional method is presented, for example,  in 
Chapters 3 and 4 of Ref.~\cite{Brown}.  We should note that  the 
tensor algebra used before the extension to $\nu \neq 3$ spatial 
dimensions that commenced in Eq.~(\ref{nunu}) was restricted to 
$\nu = 3 $. This use of the $\nu = 3$ tensor algebra is justified because, 
with the dimensional continuation method, no divergence appears in the 
$\nu \to 3$ limit.} 
method of regularization in our work --- other methods would require the 
introduction of counter terms to cancel divergent quantities.  We have
captured the $\nu \to 3$ limit
\begin{eqnarray}
\label{eqn:sdt}
\Sigma_{dt}(W) 
&=&
- \frac{g_{dt}^2}{4\pi } \, 
 \, 2m_{dt} \, i \, p_{dt} \,,
\end{eqnarray}
and
\begin{eqnarray}
\label{eqn:sna}
\Sigma_{n\alpha}(W)
&=&
- \frac{1}{3} \, \frac{g_{n\alpha}^2}{4\pi } \, 
 p_{n\alpha}^4 \, 2m_{n\alpha} \, i \,  p_{n\alpha} \,. 
\label{Lnalpha}
\end{eqnarray}

With these self-energy functions, the unstable particle Green's functions
reads
\begin{align}
   G_*^{-1}(W) &= -(W  + \epsilon_*) + i \frac{g_{dt}^2}{4\pi }
 2m_{dt} p_{dt} 
 \nonumber \\
 &+ i \frac{g_{n\alpha}^2}{4\pi } \, 
 \frac{2}{3} m_{n\alpha} \, i \,  p_{n\alpha}^5 \,. 
\label{unstableG}
\end{align}

\subsection{$d\,t \to n\,\alpha$ Reaction Amplitude and Cross Section}
\label{reactamp}

\begin{figure*}[ht]
   \includegraphics[width=7cm]{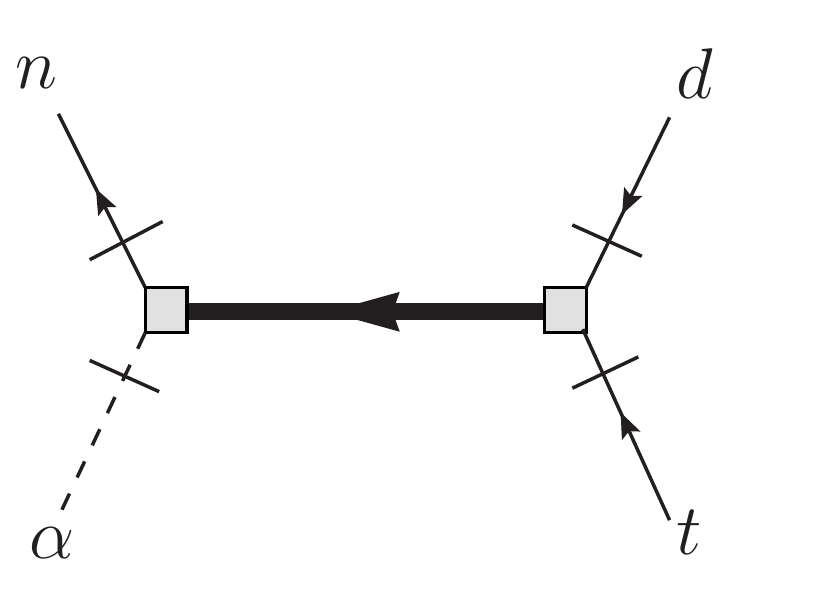}
   \caption{\label{fig:dtna}Graphical representation of the transition 
      amplitude for $dt\to n\alpha$, Eq.~\eqref{Tdtnalpha}, excluding 
      Coulomb corrections. The thick, directed line is the interacting 
      $^5$He$^*$ Green's function illustrated in Figs.~\ref{fig:5heprop} 
      and \ref{fig:dtse-nocoul}. Solid, directed lines are non-zero spin 
      particles and the dashed line is the spin-zero $\alpha$-particle. 
      Shaded boxes are $g_{n\alpha}$ and $g_{dt}$ vertices.  Hash marks 
      on the external lines indicate that they correspond to on-shell
      asymptotic particles, not propagators.}
\end{figure*}

Expanding out the interaction picture time-ordered evolution operator 
(\ref{U}) in powers of the interaction Lagrange function (\ref{int}) and
resumming the resulting bubble chains expresses the $dt \to n\alpha$
fusion amplitude as
\begin{align}
\label{structure}
T^{\quad l}_{n\alpha \,dt}({\bf p}_{n\alpha};{\bf p}_{dt}) &=
{\cal Q}^{\quad l}_{n\alpha\, dt}({\bf p}_{n\alpha}) 
\tilde T_{n\alpha \,dt}(W),
\intertext{with}
\label{Tdtnalpha}
 \tilde T_{n\alpha \,dt}(W) &= g_{n\alpha} \, G_*(W) \, g_{dt} \,,
\end{align}
which is expressed diagrammatically in Fig.~\ref{fig:dtna}.
The kinematical structure of the reaction amplitudes has the following
ingredients. The relative momentum  ${\bf p}_{n\alpha}$ 
is constrained by the energy conservation equation
\begin{equation}
\frac{{\bf p}_{n\alpha}^2}{2 m_{n\alpha}}
= W + \epsilon_n + \epsilon_\alpha \,.
\end{equation}
The tensor (\ref{calT}), together with $\sigma^k$, is sandwiched between 
the $n \, \alpha$ fields  to form the composite field 
$\Psi^{\dagger  \, l}_{\alpha \, n}({\bf r},t)$ that initiates the 
reaction via the expansion of the time-ordered interaction operator
(\ref{U}). The differential operations that appear in 
$\Psi^{\dagger  \, l}_{\alpha \, n}({\bf r},t)$  become
replaced by momenta in constructing the scattering amplitude, and these
momenta combine to produce ${\bf p}_{n\alpha}$ --- as they must
to keep the theory Galilean invariant.  The construction entails the angular
momentum $l=2$ tensor
\begin{equation}
{\cal T}^{mk}({\bf p}) = p^m p^k - 
     \frac{1}{3} \, {\bf p}^2 \, \delta^{mk} 
\label{tensor}
\end{equation}
contracted with the projection matrix $P_{3/2}^{kl}$ of
the unstable $^5{\rm He}^*$ propagator [leaving the scalar part 
$G_*(W) $ displayed in the scalar amplitude (\ref{Tdtnalpha})] to produce
the factor shown in Eq.~(\ref{structure}):
\begin{eqnarray}
{\cal Q}^{\quad l}_{n\alpha \,dt}({\bf p}_{n\alpha}) =
 \sigma^m  \, {\cal T}^{mk}({\bf p}_{n\alpha}) \, P_{3/2}^{kl} \,. 
\label{QR}
\end{eqnarray}

The total cross section involves the solid angle integration
\begin{align}
&\int \frac{d\Omega}{4\pi} \, 
{\cal Q}^{\quad k}_{n\alpha \,dt}({\bf p}_{n\alpha}) 
{\cal Q}^{\quad l}_{n\alpha \,dt}({\bf p}_{n\alpha}) 
\nonumber\\
&
=
\int \frac{d\Omega}{4\pi} P_{3/2}^{km} \,
 {\cal T}^{mn}({\bf p}_{n\alpha}) \, \sigma^n 
 \sigma^r  \, {\cal T}^{rs}({\bf p}_{n\alpha}) \, P_{3/2}^{sl}
\nonumber\\
&
= \frac{1}{15} \, \left({\bf  p}_{n\alpha}^2 \right)^2 \, 
\left[ \delta^{mr} \,\delta^{ns} 
+ \delta^{ms} \, \delta^{nr}  - \frac{2}{3} \delta^{mn} \, \delta^{rs} \right] 
\nonumber \\ &\times
    P_{3/2}^{km} \, \left[ \delta^{nr} 
      + i \epsilon^{nrq} \sigma^q \right] P_{3/2}^{sl}
\nonumber\\
&
= \frac{1}{3}\left({\bf  p}_{n\alpha}^2 \right)^2 \, P_{3/2}^{kl} \,.
\label{hard}
\end{align}
The calculation of the last line of Eq.~(\ref{hard}) from that
preceding it is facilitated by using a matrix notation and noting that
the form $\vec\sigma \cdot \vec{\cal S}$ enters, of which $P_{3/2}$ is
an eigenvector with eigenvalue $+1$. 

\begin{figure*}[ht]
   \includegraphics[height=3cm]{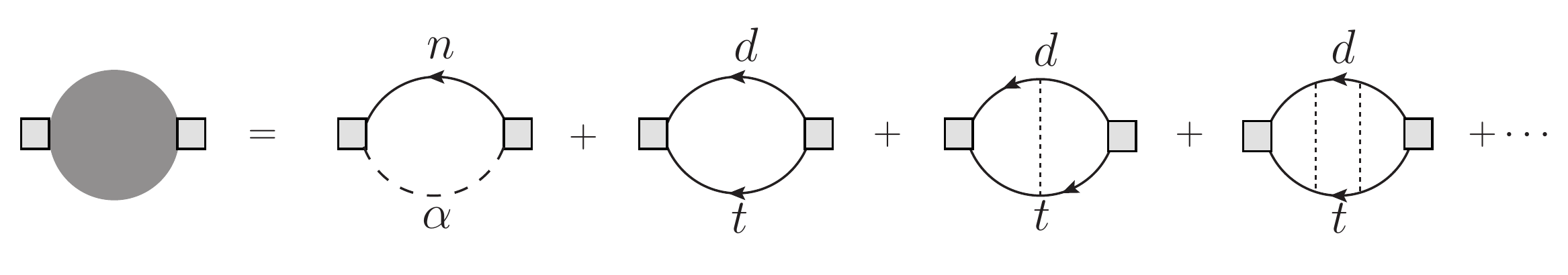}
   \caption{\label{fig:dtse-coul}Self-energy of the $^5$He$^*$, including
      the instantaneous Coulomb interaction. The first two terms are
      identical to those in Fig.~\ref{fig:dtse-nocoul}. The following,
      additional terms are corrections due to instantaneous Coulomb
      exchanges between the charged particles.  The ellipsis denote
      that the instantaneous Coulomb exchanges are summed to all
   orders, as given by Eq.~\eqref{eqn:sigcdt}.}
\end{figure*}

Thus the $dt \to n\alpha$ reaction total cross section in our
approximation that has an unstable, $3/2^+$ resonant intermediate
$^5{\rm He}$ state is given by
\begin{align}
\sigma_{dt \to n\alpha} &=
\frac{1}{6}
 \frac{m_{n\alpha}}{(2\pi)^2} \frac{p_{n\alpha}}{v_{dt}} 
\int d\Omega\, {\rm tr }
 \nonumber \\ &\times
{T^{\quad l}_{n\alpha dt}
\left( {\bf p}_{n\alpha} ; {\bf p}_{dt} \right)}^\dagger
T^{\quad l}_{n\alpha dt}\left( {\bf p}_{n\alpha} ; {\bf p}_{dt} \right),
\end{align}
where now  ${\rm tr}$ denotes the trace over the spin $1/2$ parts.
Using the result (\ref{hard}) with the trace formula $ {\rm tr} \,
P_{3/2}^{ll} = 2 (3/2) + 1 = 4 $ which simply counts the number of
spin $3/2$ states, we obtain
\begin{equation}
\sigma_{dt \to n\alpha}  =
\frac{8}{9} \, 4\pi \, 
 m_{n\alpha} \, \frac{p^5_{n\alpha}}{v_{dt}} \, \frac{g^2_{dt}}{4\pi} 
\, \frac{g^2_{n\alpha}}{4\pi} \, \left| G_*(W)\right|^2 \,.
\label{fusion}
\end{equation}
The energy dependence of the cross section is best revealed if we use
\begin{align}
   W &= \frac{p_{dt}^2}{2m_{dt}} - \epsilon_d - \epsilon_t \nonumber \\
     &= \frac{p_{dt}^2}{2m_{dt}} + Q - \epsilon_\alpha \nonumber \\
     &= \frac{p_{n\alpha}^2}{2m_{n\alpha}} - \epsilon_\alpha,
\end{align}
where, we recall, $\epsilon_n \equiv 0$ sets the energy scale, and the
energy release in the reaction $Q$ is given by the binding energy difference,
$Q = \epsilon_\alpha - \epsilon_d - \epsilon_t$. Thus
\begin{align}
   p_{n\alpha} &= \left[ 2 m_{n\alpha}  
           \left( \frac{p_{dt}^2}{2 m_{dt}} + Q \right) \right]^{1/2}
	   \nonumber \\ &
= \left( 2 m_{n\alpha} \, Q \right)^{1/2} 
\left[ 1 + \frac{p_{dt}^2}{2m_{dt} \, Q } \right]^{1/2} \,,
\end{align}
where we write the second equality to emphasize that, since in the 
energy region of interest
$p_{dt}^2 /2m_{dt} \ll Q \,,$ 
the momenta $p_{n\alpha}$ is nearly a constant determined by the 
energy release $Q$. 
Thus we write the squared unstable particle's Green's function as
\begin{align}
&\left| G_*(W)\right|^{-2}  
   \nonumber \\ 
&= \Bigg[- \left( \frac{p_{dt}^2}{2m_{dt}} - E_* \right)
    - {\rm Re}  \Sigma_{dt}(W) 
    -  {\rm Re}  \Sigma_{n\alpha}(W) \Bigg]^2
    \nonumber \\ &
  +
\Bigg[ {\rm Im}  \Sigma_{dt}(W) 
   +  {\rm Im}  \Sigma_{n\alpha}(W) \Bigg]^2 
\nonumber\\
&=
\Bigg[ \frac{p_{dt}^2}{2m_{dt}} - E_* \Bigg]^2 +
\Bigg[ \frac{g_{dt}^2}{2\pi}  m_{dt}  p_{dt} 
   + \frac{g_{n\alpha}^2}{6\pi}   m_{n\alpha}  p_{n\alpha}^5 
       \Bigg]^2.
\label{square}
\end{align}
in which we write the unrenormalized energy $\epsilon_*$ in terms of the
initial $dt$ energy as 
\begin{equation}
E_* =  \epsilon_d + \epsilon_t  -\epsilon_* \,.
\end{equation}
The first equality in Eq.~(\ref{square}) is in a form that we shall shortly
make use of when we take account of Coulomb corrections.

\section{Coulomb Corrections}
\label{CCcorr}

Including Coulomb corrections, the $dt \to n\alpha$ reaction is still
described by the diagram in Fig.~\ref{fig:dtna}, but with two
significant changes. The $dt$ entrance channel that connects these
particles to the unstable, interacting $^5$He$^*$ resonant
Green's function involves a point interaction. Hence, one effect of the
Coulomb force between the $dt$ in the fusion process is to multiply
the cross section by the square of the Coulomb wave function
$\psi^{(C)}_{{\bf p}_{dt}}(0)$ at the origin. Thus the initial shaded
box at the right in Fig.~\ref{fig:dtna} must now contain 
$\psi^{(C)}_{{\bf p}_{dt}}(0)$ multiplying the coupling constant $g_{dt}$.
The other effect of the Coulomb interactions is to modify the $dt$ loop
graphs in the $^5{\rm He}^*$ resonant Green's function by including 
arbitrary numbers of instantaneous Coulomb interactions as depicted
in Fig.~\ref{fig:dtse-coul}. These heuristic remarks are substantiated 
in Appendix \ref{coul}. Here we shall simply state and discuss the 
results of these Coulomb corrections.

The cross section involves the square of the amplitude and thus the
square of the Coulomb wave function at the origin,
\begin{equation}
\left| \psi^{(C)}_{{\bf p}_{dt}}(0) \right|^2 = 
\frac{2\pi\eta \, \exp\{-2\pi\eta \} }{ 1 - \exp\{-2\pi\eta\} } \,.
\label{psiC}
\end{equation}
This is essentially the familiar Gamow barrier penetration factor.
For our deuteron-triton system, each with a single electron
charge $e$, in ordinary cgs units,
\begin{equation}
\eta = \frac{e^2}{v_{dt}} = \frac{e^2 m_{dt}}{p_{dt}} \,.
\label{eta}
\end{equation}
It is sometimes convenient to write
\begin{equation}
\eta = \frac{1}{b_0 \, p_{dt}} \,, 
\end{equation}
with, in our units in which $\hbar = 1$, $b_0$ is the Bohr radius for the $dt$ 
system,
\begin{equation}
b_0 = \frac{1}{e^2 m_{dt}} = 24.04 \, {\rm fm} 
    = 0.1218 \, \frac{c}{\rm MeV} \,,
\end{equation}
where $1 \, {\rm fm} = 10^{-13} \, {\rm cm} \,,$ and we have made use
of $\hbar  c = 197.33 \, {\rm MeV \,\, fm }$ in writing the last equality.  
In our theory, the Coulomb corrections to the 
intermediate state nuclear interactions appear only in the unstable field 
propagator. Thus, including all the Coulomb effects, the previous fusion
cross section (\ref{fusion}) becomes
\begin{align}
   \sigma_{dt \to n\alpha} &=
\frac{8}{9} \, 4\pi \,
 m_{n\alpha} \, \frac{p_{n\alpha}^5}{v_{dt}} \,
\frac{g^2_{dt}}{4\pi} \, \frac{g^2_{n\alpha}}{4\pi}  \, 
\nonumber \\ &\times
\left| \psi^{(C)}_{{\bf p}_{dt}}(0) \right|^2 \,
\left|G_*^{(C)}(W)\right|^2 \,.
\label{fusionC}
\end{align}

Here we use an additional superscript on the unstable particle's
interacting Green's function $G_*^{(C)}(W)$ to note that it now
includes the effect of the Coulomb interaction. The only effect of
this interaction is on the previous $dt$ loop function
(\ref{dtselfE}) that contains charged particles, with
\begin{align}
   \Sigma_{dt}(W) &\to \Sigma^{(C)}_{dt}(W) \nonumber \\
&=
-i \, g_{dt}^2 \,\int_0^\infty \, dt \, e^{ i W \, t } \, 
\langle {\bf 0} , t | {\bf 0} , 0 \rangle^{(C)}_{dt \,\, \rm rel} \,,
\label{selfEC}
\end{align}
corresponding to the infinite series of diagrams indicated in
Fig.~\ref{fig:dtse-coul} containing the $dt$ intermediate state.
Introducing a complete set of incoming wave intermediate eigenstates
gives
\begin{align}
\langle {\bf 0} , t | {\bf 0} , 0 \rangle^{(C)}_{dt \,\, \rm rel} 
&=
\int \frac{(d^3{\bf p}')}{(2\pi)^3} \,
\langle {\bf 0} , t |{\bf p}' \, {\rm in} \rangle^{(C)}_{dt \,\, \rm rel} 
\langle {\bf p}' \, {\rm in} | {\bf 0} , 0 \rangle^{(C)}_{dt \,\, \rm rel} 
\nonumber\\
&=
e^{i(\epsilon_d + \epsilon_t)t} \,
\int \frac{(d^3{\bf p}')}{(2\pi)^3} \,
\exp\left\{ - i  \frac{{p'}^2}{2m_{dt}} \, t \right\} \,
\nonumber \\ &\times
\left| \psi_{p'}^{(C)}(0) \right|^2 \,,
\label{decompose}
\end{align}
in which $m_{dt}$ is the reduced mass of the $dt$ system and 
$ \psi_{p'}^{(C)}(0) $ is the Coulomb wave function (\ref{psiC}).  Using the
$dt$ relative momentum $p_{dt}$ so that the energy in the center of
mass is given by
\begin{equation}
W + \epsilon_d + \epsilon_t = \frac{p_{dt}^2}{2m_{dt}} \,,
\end{equation}
performing the time integration in Eq.~(\ref{selfEC}) with the 
decomposition (\ref{decompose}), and also performing the angular part
of the momentum integral produces
\begin{align}
\overline{\Sigma^{(C)}}_{dt}(W)
&= g_{dt}^2 \, \frac{m_{dt}}{\pi^2} \,
 \int_0^\infty dp' \, \left| \psi^{(C)}_{p'}(0) \right|^2 
 \nonumber \\ &\times
\frac{{p'}^2} {p_{dt}^2 - {p'}^2 + i 0^+ } \,.
\end{align} 
At large momenta, the Coulomb wave function approaches the free-particle
limit, $ p' \to \infty \,: \,   \psi^{(C)}_{p'}(0) \to 1 $. Hence the 
integral here does not converge at large momenta. We have noted this divergence 
by temporarily placing an overline on the function. Previously, we dealt
with this convergence problem by employing the dimensional continuation
method which simply removes this divergence in three dimensions. Here, 
however, we have an integrand involving the square of the Coulomb wave
function and the application of the dimensional continuation method is
more complex.  There is no real problem here  because the divergence produces 
an additional constant that is simply removed by an additive renormalization
of the energy $\epsilon_*$, a renormalization which we shall assume
has been implicitly performed.  Thus we simply subtract the asymptotic
limit and replace
\begin{align}
   \frac{{p'}^2} {p_{dt}^2 - {p'}^2 + i 0^+ } &\to
\frac{{p'}^2} {p_{dt}^2 - {p'}^2 + i 0^+ }  + 1 
\nonumber \\
&= \frac{p_{dt}^2} {p_{dt}^2 - {p'}^2 + i 0^+ } \,,
\end{align}
remove the overline from the self-energy function, and write
\begin{align}
\Sigma^{(C)}_{dt}(W)
&=
 g_{dt}^2 \frac{m_{dt} p_{dt}^2}{\pi^2}
 \int_0^\infty dp' \, \frac{\left| \psi^{(C)}_{p'}(0) \right|^2} 
{p_{dt}^2 - {p'}^2 + i 0^+ }.
\end{align} 

This shows explicitly that the Coulomb-corrected self-energy function 
$\Sigma^{(C)}_{dt}(W)$ is the boundary value of a function that is analytic 
in the upper half complex $ p_{dt}^2 / 2 m_{dt} $ plane.  This is because 
it is the Fourier transform of a retarded, causal response function. We 
conclude that the real part of $\Sigma^{(C)}_{dt}(W)$ must accompany its 
imaginary part to keep the proper, complete analytic function. 
For physical, real energies we may use
\begin{align}
   {\rm Im} \, \frac{1}{p_{dt}^2 - {p'}^2 + i 0^+ } 
   &= - \pi \delta(p_{dt}^2 - {p'}^2 ) 
	       \nonumber \\ &
               = - \frac{\pi}{2 p_{dt}} \,  \delta(p_{dt} - p' ) 
\end{align}
to compute the imaginary part,
\begin{eqnarray}
{\rm Im} \, \Sigma^{(C)}_{dt}(W)
&=&
-  g_{dt}^2 \, \frac{m_{dt} \, p_{dt}}{2 \pi} \,
  \left| \psi^{(C)}_{p_{dt}}(0) \right|^2 \,.
\end{eqnarray} 

To have an explicit representation of the real part, we use
\begin{equation}
\left| \psi^{(C)}_{p'}(0) \right|^2 = 
\frac{2\pi\eta' }{ \exp\{2\pi\eta'\} -1 } \,,
\end{equation}
and change the integration variable to
$
t = \eta' =  1 / (p'\, b_0) \,,
$
obtaining
\begin{align}
\Sigma^{(C)}_{dt}(W)
&=
 g_{dt}^2  \frac{2  m_{dt}}{\pi  b_0}    \int_0^\infty dt \, t 
 \frac{1}{t^2 - \eta^2 + i 0^+ }  \nonumber \\ &\times
 \frac{1}{\exp\{2\pi  t \} -1 } ,
\end{align}
in which
$
\eta =  1 / (p_{dt}\, b_0) \,.
$
This new form of the self-energy function is simply related to the $\psi$ 
function --- the logarithmic derivative of the $\Gamma$ function --- 
because of the integral representation\footnote{See, for example, 
Sec.~{\bf 1.7.2}, Eq.~(27) of Ref.~\cite{Bat}, or Sec.~8.361, Eq.~(3) 
of Ref~\cite{Grad}.} 
\begin{align}
   \psi(z) &= \ln z - \frac{1}{2z} - 2 \int_0^\infty dt \, t \,
\frac{1}{t^2 + z^2} \nonumber \\ &\times 
\frac{1}{\exp\{2\pi \, t \} -1 },
\label{rep}
\end{align}
where it is assumed that ${\rm Re}  z > 0$. Hence
\begin{eqnarray}
\label{eqn:sigcdt}
\Sigma^{(C)}_{dt}(W) = - g_{dt}^2  \frac{m_{dt}}{\pi  b_0} 
\left[ \psi(i\eta) -  \ln  \eta - \frac{\pi}{2}   i - \frac{i}{2\eta}
      \right],
\end{eqnarray}
which gives the real part
\begin{eqnarray}
{\rm Re} \, \Sigma^{(C)}(W) = - \frac{g_{dt}^2}{4\pi} \, \Delta(W) \,,
\end{eqnarray}
where
\begin{eqnarray}
\Delta(W) =
\frac{4m_{dt}}{ b_0} 
\left[ {\rm Re} \, \psi(i\eta) - \, \ln \, \eta  \right] \,,
\label{delta}
\end{eqnarray}

We pause to record some formulae that can prove to be useful checks on 
numerical computations.  It follows from the standard series 
development\footnote{See, for example, Sec.~{\bf 1.7}, Eq.~(3) of 
Ref.~\cite{Bat}.} of the $\psi$ function that
\begin{equation}
{\rm Re} \, \psi(i\eta) = - \gamma + \sum_{k=1}^\infty \frac{1}{k}
                                           \frac{\eta^2}{k^2 + \eta^2} \,,
\end{equation}
which gives the small $\eta$ behavior
\begin{align}
   {\rm Re} \, \psi(i\eta) &= - \gamma + \zeta(3) \, \eta^2 + \dots 
	      \nonumber \\ &
                        \simeq - 0.5772157 + 1.20206 \, \eta^2 \,.
\end{align}
Moreover,
\begin{equation}
{\rm Re} \, \psi(i\eta) = - \gamma + \frac{\eta^2}{1 + \eta^2} 
           + \frac{1}{2} \frac{\eta^2}{4 + \eta^2} + R(\eta) \,,
\end{equation}
where we have the bound
\begin{align}
\left| R(\eta) \right| \leq B(\eta) &=
\eta^2 \, \sum_{k=3}^\infty \frac{1}{k^3} 
= 
\eta^2 \, \left[ \zeta(3) - 1 - \frac{1}{8} \right]
\nonumber \\ &
= \eta^2 \left[ 0.07706  \right] \,.
\end{align}
The large $\eta$ limit\footnote{See, for example, Sec.~{\bf 1.18}, 
Eq.~(7) of Ref.~\cite{Bat}.}  gives
\begin{equation}
{\rm Re} \, \psi(i\eta) - \ln \eta = \frac{1}{12 \, \eta^2} 
- \frac{1}{120} \, \frac{1}{\eta^4} + \dots \,.
\end{equation}

The inverse unstable field Green's function now reads 
\begin{equation}
G_*^{(C) \, -1}(W) = -(W + \epsilon_*) - \Sigma^{(C)}_{dt}(W) 
               - \Sigma_{n\alpha}(W) \,,
\end{equation}
with $\Sigma^{(C)}_{dt}(W)$ the function that we have just computed,
while
\begin{eqnarray}
\Sigma_{n\alpha}(W)&=&
 - i \, \frac{2}{3} \, \frac{g_{n\alpha}^2}{4\pi} \,
                           m_{n\alpha}  p_{n\alpha}^5
\end{eqnarray}
is as before.

The absolute square of the unstable field Green's function 
that is needed for the fusion cross section (\ref{fusionC}) is produced 
by the trivial notational change $\Sigma_{dt} \to \Sigma^{(C)}_{dt}$ in the
formula (\ref{square}), producing
\begin{align}
\label{theend}
&\left| G_*^{(C)}(W)\right|^{-2}  
\nonumber \\ 
&=
\Bigg[ \frac{p_{dt}^2}{2m_{dt}} - E_*
    - \frac{g_{dt}^2}{4\pi} \Delta(W) \Bigg]^2
\nonumber\\
 & +
\Bigg[ \frac{g_{dt}^2}{2\pi}  m_{dt}
   p_{dt} \left| \psi^{(C)}_{p_{dt}}(0) \right|^2
+ \frac{g_{n\alpha}^2}{6\pi} m_{n\alpha} p_{n\alpha}^5 \Bigg]^2 
\end{align}
in which we again write the unrenormalized energy of the unstable particle 
in terms of the initial $dt$ energy as 
\begin{equation}
E_* =  \epsilon_d + \epsilon_t - \epsilon_* \,.
\end{equation}

\section{Conclusion and Discussion}
\label{sec:conc}

We have demonstrated that an excellent description of the $dt\to
n\alpha$ reaction in the resonance region is obtained with an
effective quantum field theory that entails only the interaction of
the initial and final particles $dt$ and $n\alpha$ with an unstable
$^5$He$^*$ $3/2^+$ unstable field. The fit, with a $\chi^2$ per
degree-of-freedom less than unity, is achieved with just three
parameters: the energy of the $^5$He$^*$ resonance $E_*$, and the two
coupling parameters $g_{dt}$ and $g_{n\alpha}$. We have calculated the
Coulomb corrections exactly in Sec.~\ref{CCcorr} and taken into
account their effect on both the incoming $dt$ particles and on the
strong interactions which transmute these particles into the final
$n\alpha$ particles.

It is worthwhile noting that, early on, we found that the $dt\to
n\alpha$ resonance could not be fit by using an intermediate $3/2^+$
unstable field which had the right sign free-particle Lagrangian. The
fit was so poor that a $\chi^2$ value to describe it is not
meaningful. Roughly, if the parameters were adjusted to fit the
maximum of the cross section resonance, then the data around half
maximum were about 30\% above the fit.  This result led us to add an
additional contact interaction that coupled the initial and final
particles in a $3/2^+$ state. We chose this form of the contact
interaction so that it would enter into the self-energy function of
the $3/2^+$ field's Green's function and hence would be a candidate to
alter the resonance shape produced by the theory. As described in
Eq.~(\ref{ugh}), the final $n\alpha$ particles are produced in a
$3/2^+$ state by the field combination $\bm{\Psi}_{\alpha \, n}$ and
hence the contact interaction was chosen to have the form 
\begin{equation} 
\lambda \left[ \bm{\Psi}^\dagger_{\alpha \, n} \cdot
\bm{\phi}_d \,\, \psi_t + \psi^\dagger_t \,\, \bm{\phi}^\dagger_d
\cdot \bm{\Psi}_{\alpha \, n} \right] \,.  
\end{equation} 
As should be expected, the introduction of the additional free
parameter $\lambda$ improved the fit. However, the improvement was not
significant.  With the parameters again chosen to have the resonance
maximum fitted, the data around half maximum were now about 20\% above
the fit. Therefore, we reverted to the simple, previous single
interaction with an intermediate unstable field, and changed the sign
of its free-particle propagation to achieve the excellent description
of the $dt$ fusion reaction that is presented in this paper. The quality 
of this fit, about 1\% deviation for most of the data points, was a dramatic 
improvement over the other work just discussed.

The relationship between the effective field theory applied here and
the $R$-matrix approach is presented in the following paper
\cite{BHP}.  It establishes an identity between the effective field
theory approach and that of the $R$ matrix, in the limit that the
$R$-matrix channel radii go to zero.  In this limit, the $R$-matrix
parameterization provides an excellent fit of the data when generalized
to allow for ``unphysical'' negative values of the reduced widths.
These unphysical couplings are directly related to the
wrong-sign free-particle Lagrange function used in the present
work. This is a promising indication that carrying out a multichannel,
many-resonance effective field theory description of light nuclear
reaction data is possible by suitably generalizing current $R$-matrix
methods and codes that are already in use.

\acknowledgments

We are indebted to Mark Paris for his detailed comments that have
improved the presentation of this paper.  This work was carried out
under the auspices of the National Nuclear Security Administration.

\appendix

\section{Galilean Invariance}
\label{galinv}

The consequences of translational and rotational invariance are obvious.
The consequences of Galilean invariance --- the invariance of the 
theory under `boosts' to a moving coordinate frame --- are less
obvious, although perhaps they are clear once they have been 
derived. Here we sketch out the derivation of  some of these 
consequences.\footnote{A detailed exposition of the Galilean
invariance of a non-relativistic field theory
is presented, for example, in the discussion of  Problem 1 on Page 118 of 
ref. \cite{Brown}. The explicit solution
of this problem has been done in the 
MIT 8.323 Second homework solutions by J. Goldstone, February, 1995,
but this may not be readily accessible.}

We denote the generator of boosts to moving frames by $ {\bf G} $.
It has the construction
\begin{equation}
{\bf G} = {\bf P} \, t - M \, {\bf R}(t) \,,
\label{G}
\end{equation}
where, for a general set $\{\chi_a, \chi^\dagger_a \} $ of 
non-relativistic fields, 
\begin{equation}
{\bf P} = {\sum}_a \int (d^3{\bf r}) \, \chi_a^\dagger({\bf r},t)
          \,  \frac{1}{i} \nabla \,  \chi_a({\bf r},t)
\end{equation}
is the total momentum operator of the system, and
\begin{equation}
M \, {\bf R}(t) =  {\sum}_a \int (d^3{\bf r}) \, \chi_a^\dagger({\bf r},t)
          \, m_a \, {\bf r} \,  \chi_a({\bf r},t)
\end{equation} 
defines a the center-of-mass operator ${\bf R}$, with $m_a$ the 
kinematical mass of the particle, bound state, or resonance created 
and annihilated by the $a$-th field and $M = {\sum}_a \, m_a $ is
the total mass of all these (perhaps quasi-) particles.   
The kinematical mass $m_a$ is the mass that appears in the kinetic
energy part of the Hamiltonian density 
$\chi^\dagger_a (- \nabla^2 / 2 m_a) \chi_a$ 
for the $a$-th field.

The continued iteration of the infinitesimal Galilean boosts yields
the unitary transformation
\begin{equation}
U({\bf v}) = \exp\left\{ i \, {\bf G} \cdot {\bf v} \right\}
= \exp\left\{ i \left[ {\bf P} \, t - M \, {\bf R}(t) \right]
            \cdot {\bf v} \right\} 
\end{equation}
to a frame moving with the finite velocity ${\bf v}$. Standard 
methods now show that the response of a field $\chi_a$ to a 
finite Galilean boost is given by
%5
\begin{align}
\chi_a({\bf r},t) &\to
 U^{-1}({\bf v})  \chi_a({\bf r},t)  U({\bf v}) 
\nonumber\\
&=
\exp\left\{- i  m_a   \left[ {\bf r} \cdot {\bf v}
+ \frac{1}{2}  {\bf v}^2  t \right] \right\}  
\nonumber \\ &\times
\chi_a({\bf r}+ {\bf v} t ,t) .
\label{Gchi}
\end{align}
We shall make use of three implications of this transformation.

Nucleons can be put into bound or resonant states. Hence the effective
field theory can contain interactions of the schematic form 
\begin{equation}
H_{int} \sim 
\int (d^3{\bf r}) \, \chi_{b_n}^\dagger \cdots \chi_{b_1}^\dagger
       \,  \chi_{a_m} \cdots \chi_{a_1} \,.
\label{genint}
\end{equation}
In view of the response of the individual fields given by Eq.~(\ref{Gchi}), 
Galilean invariance  requires that
\begin{equation}
m_{b_n} + \dots + m_{b_1} = m_{a_m} + \dots + m_{a_1} \,.
\label{massconst}
\end{equation}
Thus, for example, the inertial mass of the deuteron is the sum of the
neutron and proton masses, $m_d = m_n + m_p$, the triton mass is
$m_t = 2 m_n + m_p $, and the alpha mass is $m_\alpha = 2 m_n + 2m_p$.

The second consequence of Galilean invariance that we mention is that
a generic free-particle Lagrangian must be of the form
\begin{align}
   L_\smA &= \int (d^3{\bf r}) 
 \chi_\smA^\dagger({\bf r},t) \nonumber \\ 
 &\times \left[ i \frac{\partial}{\partial t} 
 - \frac{1}{2 m_\smA} \, \left( \frac{1}{i} \nabla \right)^2 
       - \epsilon_\smA \right] \chi_\smA({\bf r} , t) \,.
\end{align}
Galilean invariance requires that the relative signs of the time
and spatial derivatives must appear as they are shown here.

The final application that we need involves field derivatives. 
To obtain Galilean invariant interactions, we need field combinations
with derivatives that undergo a simple phase change under Galilean
transformations. This does not happen with a single derivative of a
single field. However, with a pair of fields $\chi_a$, $ \chi_b$, we 
can define a derivative operation
\begin{equation}
{\cal P}^k_{ba} =
\frac{m_{ab}}{m_a} \frac{1}{i}\stackrel{\rightarrow}{\nabla}^k -
\frac{m_{ab}}{m_b} \frac{1}{i} \stackrel{\leftarrow}{\nabla}^k \,,
\label{calP}
\end{equation}
in which $\rightarrow$ calls for the derivative to act to the right,
and  $\leftarrow$ calls for the derivative to act to the left. 
Here
\begin{equation}
m_{ab} = \frac{m_a \, m_b}{M_{ab}} \,, 
\end{equation}
is the reduced mass of the $a,b$ pair with 
\begin{equation}
       M_{ab} = m_a + m_b
\end{equation}
the total mass. It is now easy to see that
\begin{align}
&U^{-1}({\bf v}) \chi_b({\bf r},t) {\cal P}^k_{ba} \chi_a({\bf r},t) 
U({\bf v})=
\nonumber\\
& e^{- i \, (m_b + m_a) \, [ {\bf r} \cdot {\bf v} +
         \frac{1}{2} \, {\bf v}^2 \, t ] }  
\chi_b({\bf r}+ {\bf v} t ,t) 
\nonumber \\ &\times {\cal P}^k_{ba}
\chi_a({\bf r} + {\bf v} t ,t),
\label{Gchii}
\end{align}
which is the desired transformation law. 

\section{Spin Structure}
\label{spin}

Here we shall describe some algebraic properties of the spin matrices
that are needed in the text. We shall keep to our $\hbar = 1$ convention
so as to simplify the notation.  The 
action on any field $\chi$ of an infinitesimal rotation generated by
the field operator angular momentum ${\bf J}$ is given by 
\begin{equation}
\left\{{\bf L} + {\bf S} \right\} \chi({\bf r},t) =
       \left[\chi({\bf r},t) , {\bf J} \right] \,,
\end{equation}
in which
\begin{equation}
{\bf L} = {\bf r} \times \frac{1}{i} \nabla
\end{equation}
are the orbital angular momentum differential operators and ${\bf S}$
are the spin matrices.  The structure of the rotation  group is
conveyed by
\begin{equation}
\left[ J^k , J^l \right] = i \epsilon^{klm} J^m \,,
\end{equation}
where $\epsilon^{klm}$ is the completely antisymmetrical numerical
tensor associated with vector cross product, with 
$\epsilon^{123} = 1$. This numerical tensor 
satisfies the ``double cross'' relation
\begin{equation}
\epsilon^{klm} \epsilon^{kpq} = \delta^{lp}\delta^{mq} -
                                 \delta^{lq}\delta^{mp} \,.
\label{double} 
\end{equation}
Since $L^k$ acts on the coordinates of the field while $S^k$ acts on
the (notationally suppressed) components of the fields, $L^k$ and
$S^k$  commute amongst each other, while
\begin{equation}
\left[ L^k , L^l \right] = i \epsilon^{klm} L^m \,,
\end{equation}
and
\begin{equation}
\left[ S^k , S^l \right] = i \epsilon^{klm} S^m \,.
\label{Scom}
\end{equation}
We turn now to examine the spin matrices ${\bf S}$ for spin $s$ with
$s = 1/2 \,,\, 1 \,,\, 3/2$. These matrices must obey the 
commutator (\ref{Scom}) and have the square
\begin{equation}
{\bf S}^2 = s (s + 1) \,.
\end{equation}

\subsubsection{Spin $1/2$}

For spin one-half, 
\begin{equation}
\bm{S} = \frac{1}{2} \bm{\sigma} \,,
\end{equation}
in which $\sigma^k$ are the familiar Hermitian Pauli matrices that 
obey
\begin{equation}
\sigma^k \sigma^l = \delta^{kl} + i \epsilon^{klm} \sigma^m \,. 
\label{Pauli}
\end{equation}
The antisymmetrical part of this constraint implies that the
spin $1/2$ matrices $S^k$ satisfy the angular momentum 
commutation relation (\ref{Scom}) while setting $k=l$ and
summing over these identified indices from 1 to 3 shows that
\begin{equation}
{\bf S}^2 = \frac{3}{4} 
          = \frac{1}{2} \left( \frac{1}{2} + 1 \right) \,,
\end{equation}
which identifies the spin value $1/2$.

\subsubsection{Spin $1$}

We write the spin matrices for spin one as
\begin{equation}
\bm{ S} = \bm {{\cal S}} \,.
\end{equation}
The latter are given by
\begin{equation}
\left( {\cal S}^k \right)^{lm} = i \epsilon^{lkm} \,,
\end{equation}
since one can easily verify from the relation (\ref{double}) that 
the angular momentum commutation relation (\ref{Scom}) holds and that 
\begin{equation}
\bm{{\cal S}}^2 = 1 \, (1 +1) = 2
\end{equation}

\subsubsection{Spin $3/2$}

The combination of spin $1/2$ and spin $1$ is described by the
spin matrix
\begin{equation}
\bm{{S}} = \frac{1}{2} \bm{{\sigma}} + \bm{{\cal S}} \,.
\end{equation}
This matrix obviously obeys the angular momentum commutator
(\ref{Scom}). But we need a constraint to keep to the spin
$3/2$ subspace. To obtain this constraint, we note that the
Pauli result (\ref{Pauli}) and the properties of the spin-one
matrix noted above imply the characteristic equation
\begin{equation}
\left( \bm{\sigma} \cdot \bm{{\cal S}} \right)^2 
+ \left( \bm{\sigma} \cdot \bm{{\cal S}} \right) - 2 = 0 \,.
\label{char}
\end{equation}
The eigenvalues of the matrix, which we denote by a prime,
must obey this characteristic matrix equation, and hence they are given by
\begin{equation}
\left( \bm{\sigma} \cdot \bm{{\cal S}} \right)' =
\left\{
\begin{array}{c}
-2 \,,\\
+ 1 \,.
\end{array}
\right.
\end{equation}
Therefore we have
\begin{align}
{\bf S}^2 &= \frac{3}{4} + 2 + \bm{\sigma} \cdot \bm{{\cal S}}
   \nonumber \\ &
          = 
\left\{
\begin{array}{cc}
\frac{1}{2} \left( \frac{1}{2} + 1 \right), & {\rm for} \, 
\left( \bm{\sigma} \cdot \bm{{\cal S}} \right)' = -2 ,
\\
&
\\
\frac{3}{2} \left( \frac{3}{2} + 1 \right), & {\rm for} \, 
\left(\bm{\sigma} \cdot \bm{{\cal S}} \right)' = \hphantom{-}1 ;
\end{array}
\right.
\label{sigmas}
\end{align}
the eigenvalues $-2$ and $1$ of the matrix
$ \bm{\sigma} \cdot \bm{{\cal S}} $ correspond to spins $s = 1/2$ 
and $s = 3/2$.

We need the matrix $P_{3/2}$ that projects
into the $ s = 3/2 $ subspace.  A little computation utilizing the
characteristic equation (\ref{char}) shows that  
\begin{equation}
P_{3/2} = \frac{1}{3} \left\{ \bm{\sigma} \cdot \bm{{\cal S}} + 2 \right\} 
\label{P3/2}
\end{equation}
obeys
\begin{equation}
\left( \bm{\sigma} \cdot \bm{{\cal S}} \right) P_{3/2} = P_{3/2}
= P_{3/2} \left( \bm{\sigma} \cdot \bm{{\cal S}} \right)  \,,
\label{PP3/2}
\end{equation}
and
\begin{equation}
P_{3/2}^{\,\,\,2} = P_{3/2} \,,
\end{equation}
so that it is indeed the correct projection matrix.
Writing out the components gives
\begin{equation}
P_{3/2}^{lm} = \frac{1}{3} \left\{ \sigma^k \, i \, \epsilon^{lkm} + 
                   2  \, \delta^{lm} \right\} \,.
\label{PPP3/2}
\end{equation}

In view of the result (\ref{sigmas}), a spin $3/2$ field $\psi_{3/2}$ 
must obey the constraint
\begin{equation}
\bm{\sigma} \cdot \bm{{\cal S}} \, \psi_{3/2} = \psi_{3/2} \,.
\end{equation}
and if this constraint is obeyed, the field contains only spin
$3/2$.  To obtain an equivalent constraint that is algebraically
simpler, we note that
\begin{eqnarray}
\sigma^l \, \left( \bm{\sigma} \cdot \bm{{\cal S}} \right)^{lm}
&=& - \epsilon^{lkn} \, \epsilon^{lkm} \, \sigma^n 
= - 2 \, \sigma^m \,.
\end{eqnarray}
Hence, in view of Eq.~(\ref{PP3/2}),
\begin{equation}
\sigma^l \, \left( \bm{\sigma} \cdot \bm{{\cal S}} \right)^{lm}
P_{3/2}^{mn} = - 2 \sigma^m \, P_{3/2}^{nm} = 
\sigma^m \, P_{3/2}^{nm}  \,;
\end{equation}
whence
\begin{equation}
\sigma^l \, P_{3/2}^{lm} = 0 \,,
\end{equation}
and similarly,
\begin{equation}
P_{3/2}^{lm} \,\, \sigma^m  = 0 \,.
\end{equation}
Therefore, 
\begin{equation}
\bm{\sigma} \cdot \bm{\psi}_{3/2}({\bf r},t) = 0 =
\bm{\psi}_{3/2}^\dagger({\bf r},t) \cdot \bm{\sigma} \,,
\label{3/2}
\end{equation}
constrain $\bm{\psi}_{3/2}$ and $\bm{\psi}_{3/2}^\dagger$ to 
contain only spin $3/2$.

In the text, we  need the spin $3/2$ combination of $n$,
$\alpha$ fields that couple to the ${3/2}^+$ $ ^5{\rm He}^*$ 
resonant state.  The composition of the ${3/2}^+$ state from the 
${1/2}^+$ neutron and $0^+$ alpha requires that the parity of the 
relative orbital state of the $n$ , $\alpha$ pair be even,
which is to say that the orbital angular momentum $l$ be an 
even integer, with $ l  = 2 $, the only possibility in virtue of
the rules of angular momentum addition. The differential operation  
\begin{equation}
{\cal T}^{kl}_{\alpha n} = {\cal P}^k_{\alpha n} {\cal P}^l_{\alpha n}
-\frac{1}{3} \, \delta^{kl} \, {\cal P}^m_{\alpha n} {\cal P}^m_{\alpha n} 
\end{equation}
transforms as $l=2$.  Here, ${\cal P}^k_{\alpha n}$ is defined by 
Eq.~(\ref{calP}) and a proof akin to that with Eq.~(\ref{Gchii}) shows that
\begin{equation}
\Psi^l_{\alpha n}({\bf r},t) = 
\phi_\alpha({\bf r},t) \, {\cal T}^{lm}_{\alpha n} \,
\sigma^m \, \psi_n({\bf r},t) 
\end{equation} 
has the correct Galilean transformation law. 
Using the Pauli spin formula (\ref{Pauli}) and the symmetry
$ {\cal T}^{lk}_{\alpha n}  = +{\cal T}^{kl}_{\alpha n} $ 
we compute 
\begin{equation}
\sigma^k  \Psi^k_{\alpha n}({\bf r},t) =
\phi_\alpha({\bf r},t) \, \delta^{kl} {\cal T}^{kl}_{\alpha n} \,
     \psi_n({\bf r},t) = 0 \,. 
\end{equation}
Hence, in view of Eq.~(\ref{3/2}) and the discussion following it,
we conclude that $ \Psi^k({\bf r},t) $ does indeed contain only spin
$3/2$.

\section{Coulomb corrections}
\label{coul}

To place our work in context, we first briefly review the case in
which there are only strong interactions with no Coulomb corrections. 
The four-point $n\alpha \, dt$  Green's function is the vacuum 
expectation value of a time-ordered product,
\begin{align}
&G_{n\alpha \, dt}(x'_n,x'_\alpha ; x_d , x_t) \nonumber \\
&= i^2 \left\langle 0 \left|
         \left(\psi_n(x'_n) \phi_\alpha(x'_\alpha) 
		\phi_d^\dag(x_d)\psi^\dag_t(x_t)\right)_+
			\right| 0 \right\rangle \,,
\end{align}
in which the space time coordinate $x$ is a short-hand notation for
${\bf r},t$. It has the decomposition
\begin{widetext}
\begin{align}
   G_{n\alpha \, dt}(x'_n,x'_\alpha ; x_d , x_t) &=
\int (d^4\bar{x}_n) (d^4\bar{x}_\alpha) (d^4\bar{x}_d) (d^4\bar{x}_t) \,
G_n(x'_n - \bar{x}_n) \, G_\alpha(x_\alpha'-\bar{x}_\alpha)
 \nonumber\\
&\times
\Gamma_{n\alpha \, dt}(\bar{x}_n , \bar{x}_\alpha , \bar{x}_d, \bar{x}_t)
\, G_d(\bar{x}_d - x_d) \, G_t(\bar{x}_t - x_t) \,.
\label{reduced}
\end{align}
\end{widetext}
Here $ \Gamma_{n\alpha \, dt}(\bar{x}_n , \bar{x}_\alpha , \bar{x}_d,
\bar{x}_t) $ contains only connected graphs. In our non-relativistic
theory, the single-particle propagators contain no self-energy
corrections and thus have the generic, form of the time retarded
functions
\begin{align}
G(x - x') &= \theta\left( t - t' \right)  \int \frac{(d^3{\bf p})}{(2\pi)^3}
             e^{ i {\bf p} \cdot ( {\bf r} - {\bf r}' )
	     - i E({\bf p}) ( t - t' ) }
\nonumber\\
          &= \theta\left( t - t' \right)  
   \langle {\bf r} , t | {\bf p}', t' \rangle^{(0)},
\end{align}
where in the second line we denote the free-particle
quantum-mechanical transformation function by the superscript $(0)$.
The reaction amplitude involves the asymptotic early and late time
limits $t_d = t_t \to -\infty$ and $t_n = t_\alpha  \to + \infty$ and
the identification of the propagator momenta ${\bf p}$ with the
initial momenta, ${\bf p}_d$ and ${\bf p}_t$ and final momenta, 
${\bf p}_n$ and ${\bf p}_\alpha$,
by the appropriate Fourier transformation. Therefore, aside
from conventional factors and overall delta functions of energy and
momentum conservation, the reaction amplitude $T_{n\alpha \, dt}({\bf
p}_n, {\bf p}_\alpha ; {\bf p}_d , {\bf p}_t )$ involves a Fourier
transform\footnote{Here we are sketching the non-relativistic 
analog of the relativistic reduction formula that is discussed, 
for example, in Section 2 of Chapter 6 of Brown~\cite{Brown}.}  of $ \Gamma_{n\alpha \, dt}(\bar{x}_n , \bar{x}_\alpha ,
\bar{x}_d, \bar{x}_t) $ .

We now consider effect that Coulomb corrections have on the charged
initial $dt$ state. When Coulomb interactions are present, there are
instantaneous Coulomb exchanges in the propagation of the initial $dt$
particles before strong interactions operate. Thus the product of 
the two initial
free-particle propagators $ \, G_d(\bar{x}_d - x_d) \, G_t(\bar{x}_t -
x_t) $ in Eq.~(\ref{reduced} must be replaced by the four-point $dt$
Green's function that includes the Coulomb interaction. The general
reaction amplitude involves the identification of the two initial
times, $t_d = t_t$. Our theory, in which the strong interactions are
represented by the unstable, s-channel intermediate field
$\psi_*^\dagger$ that has a local coupling to the $d$ and $t$ fields,
forces the identification $ \bar x_d = \bar x_t$ or $\bar{\bf r}_d =
\bar{\bf r}_t$ and $\bar t_d = \bar t_t$, with $\bar t_d > t_d$.
Hence the required four-point Coulomb Green's appears in the
restricted form 
\begin{align}
& -  G^{(C)}_{dt}({\bf r}_d , \bar t_d , {\bf r}_t = {\bf r}_d, 
      \bar t_t = \bar t_d ; {\bf r}_d, t_d , {\bf r}_t , t_t = t_t)
\nonumber\\
&
= \left\langle 0 \left| \psi_d(\bar{\bf r}_d, \bar t) \, 
     \psi_t({\bf r}_d ,t_d) \, \psi_t^\dagger({\bf r}_t,t_d) \, 
       \psi_d^\dagger({\bf r}_d , t_d)  \right| 0 \right\rangle^{(C)}
\nonumber\\
&
= \langle \bar{\bf r}_d = \bar{\bf r}_t ,\bar t_d | {\bf r}_d , {\bf r}_t 
                          , t_d \rangle^{(C)} \,.
\end{align}
The second line here follows from the usual non-relativistic theory where
a field operator $\psi^\dagger$ acting to the right adds a single particle 
to the state, and an operator $\psi$ acting to the left adds a particle. 
We have noted that the transition amplitude involves taking the spatial 
Fourier transform. 

Hence, passing to the transition amplitude involves
$ \langle \bar{\bf r}_d = \bar{\bf r}_t ,\bar t_d 
| {\bf p}_d , {\bf p}_t , t_d \rangle^{(C)}$.
On introducing relative and center-of-mass coordinates, this becomes
\begin{align}
   &\langle \bar{\bf r}_d = \bar{\bf r}_t ,\bar t_d 
| {\bf p}_d , {\bf p}_t , t_d \rangle^{(C)}
\nonumber \\ &
= \langle {\bf r}_{\rm rel} = 0 | {\bf p}_{dt} \rangle^{(C)}
        \langle {\bf r}_d | {\bf P}_{dt} \rangle^{(0)} 
              e^{ - i E (\bar t_d - t_d) }.
\label{CCC}
\end{align}
Here ${\bf p}_{dt}$ is the relative momentum of the $dt$ (which is the same
notation as used in the text), ${\bf P}_{dt}$ is the total momentum of this
pair of particles, and $E$ is the total energy in the initial state. Now
\begin{equation}
\langle {\bf r}_{\rm rel} = 0 | {\bf p}_{dt} \rangle^{(C)}
  = \psi^{(C)}_{{\bf p}_{dt}}(0)
\end{equation}
is precisely the Coulomb wave function at the origin introduced in the text.
Only this factor alters the Coulomb transformation function (\ref{CCC}) from
that of a free particle: 
\begin{align}
   &\langle \bar{\bf r}_d = \bar{\bf r}_t ,\bar t_d 
| {\bf p}_d , {\bf p}_t , t_d \rangle^{(C)}
\nonumber \\ &
= \psi^{(C)}_{{\bf p}_{dt}}(0) \, 
\langle \bar{\bf r}_d = \bar{\bf r}_t ,\bar t_d 
| {\bf p}_d , {\bf p}_t , t_d \rangle^{(0)} \,.
\end{align}
Hence, the `reduction formula' for the transition amplitude with Coulomb as 
well as strong interactions is the same as that for the amplitude with only 
strong interactions and no Coulomb corrections except for the overall 
multiplication of the $\psi^{(C)}_{{\bf p}_{dt}}(0)$ factor corresponding to
initial Coulomb interactions previous to the first strong interaction. Of course,
there are additional internal Coulomb corrections to the strong interactions. 
For our theory, these appear as multiple instantaneous Coulomb exchanges 
between the $d$ and the $t$ in the $dt$ loop  that contributes to the 
unstable $^5$He$^*$ Green's function as shown in Fig.~\ref{fig:dtse-coul}.

\end{document}